\documentclass[10pt,journal,compsoc]{IEEEtran}
\newcommand{\Revision}[1]{\textcolor{black}{#1}}
\newcommand{\SecRevision}[1]{\textcolor{black}{#1}}

\ifCLASSOPTIONcompsoc
  \usepackage[nocompress]{cite}
\else
  \usepackage{cite}
\fi

\ifCLASSINFOpdf
\else
\fi

\usepackage{amsthm,amsmath,amssymb}
\usepackage{mathrsfs}
\usepackage{array}
\usepackage{graphicx}
\usepackage{epsfig}
\usepackage{subfigure}
\usepackage{xcolor}
\usepackage{booktabs}
\usepackage{multirow}
\usepackage{url}
\usepackage{threeparttable}
\usepackage{enumitem}
\usepackage{algorithm}
\usepackage{algorithmic}
\begin{document}

% \title{Embracing User Diversity to Improve Generalization of Learning-Based 360$^\circ$ Video Streaming}
\title{MANSY: Generalizing Neural Adaptive Immersive Video Streaming with Ensemble and Representation Learning}

\author{Duo~Wu,
        Panlong~Wu,
        Miao~Zhang,~\IEEEmembership{Student~Member,~IEEE},
        and~Fangxin~Wang,~\IEEEmembership{Member,~IEEE}% <-this % stops a space
\thanks{Manuscript received xxx; revised xxx. This work was supported in part by NSFC with Grant No. 62293482, the Basic Research Project No. HZQB-KCZYZ-2021067 of Hetao Shenzhen-HK S\&T Cooperation Zone, NSFC with Grant No. 62471423, the Shenzhen Science and Technology Program with Grant No. JCYJ20230807114204010, the Shenzhen Outstanding Talents Training Fund 202002, the Guangdong Research Projects No. 2019CX01X104, the Young Elite Scientists Sponsorship Program of CAST (Grant No. 2022QNRC001), the Guangdong Provincial Key Laboratory of Future Networks of Intelligence (Grant No. 2022B1212010001) and the Shenzhen Key Laboratory of Big Data and Artificial Intelligence (Grant No. ZDSYS201707251409055). Fangxin Wang is the corresponding author.

Duo Wu and Panlong Wu are with Shenzhen Future Network of Intelligence Institute (FNii-Shenzhen) and School of Science and Engineering (SSE), The Chinese University of Hong Kong, Shenzhen, China. Duo Wu is also with the Shenzhen International Graduate School, Tsinghua University, China (email: wu-d24@mails.tsinghua.edu.cn;panlongwu@link.cuhk.edu.cn). 

Miao Zhang is with the School of Computing Science, Simon Fraser University, BC, Canada (email: mza94@sfu.ca).

Fangxin Wang is with School of Science and Engineering (SSE), Shenzhen Future Network of Intelligence Institute (FNii-Shenzhen), and Guangdong Provincial Key Laboratory of Future Networks of Intelligence, The Chinese University of Hong Kong, Shenzhen, China (email: wangfangxin@cuhk.edu.cn).
}}

\markboth{IEEE Transcations on Mobile Computing}%
{Wu \MakeLowercase{\textit{et al.}}: MANSY: Generalizing Neural Adaptive Immersive Video Streaming with Ensemble and Representation Learning}



\IEEEtitleabstractindextext{%
\begin{abstract}
  The popularity of immersive videos has prompted extensive research into neural adaptive tile-based streaming to optimize video transmission over networks with limited bandwidth. 
  However, the diversity of users' viewing patterns and Quality of Experience (QoE) preferences has not been fully addressed yet by existing neural adaptive approaches for viewport prediction and bitrate selection. 
  Their performance can significantly deteriorate when users' actual viewing patterns and QoE preferences differ considerably from those observed during the training phase, resulting in poor generalization. 
  In this paper, we propose \texttt{MANSY}, a novel streaming system that embraces user diversity to improve generalization. 
  % Specifically, we design an efficient multi-viewport trajectory input output architecture for viewport prediction based on ensemble learning to generalize the prediction model to accommodate users' diverse viewing patterns. We also leverage Transformer to mine users' viewing patterns to further improve the predictive performance. 
  Specifically, to accommodate users' diverse viewing patterns, we design a Transformer-based viewport prediction model with an efficient multi-viewport trajectory input output architecture based on implicit ensemble learning. 
  Besides, we for the first time combine the advanced representation learning and deep reinforcement learning to train the bitrate selection model to maximize diverse QoE objectives, enabling the model to generalize across users with diverse preferences. Extensive experiments demonstrate that \texttt{MANSY} outperforms state-of-the-art approaches in viewport prediction accuracy and QoE improvement on both trained and unseen viewing patterns and QoE preferences, achieving better generalization.
\end{abstract}

}

\maketitle

% is not selected <OR> if conference mode is selected - because compsoc
% conference papers position the abstract like regular (non-compsoc)
\IEEEdisplaynontitleabstractindextext

% \fi
\IEEEpeerreviewmaketitle

\ifCLASSOPTIONcompsoc
\IEEEraisesectionheading{\section{Introduction}\label{sec:introduction}}
\else
\section{Introduction}
\label{sec:introduction}
\fi

\IEEEPARstart{W}{ith} the rapid advancement of Virtual Reality (VR) technologies, immersive videos have attracted great attention because of the immersive experience they bring. Recent statistical report~\cite{vr_market_thomas} projects that the global installed base of VR headsets will exceed 34 million by 2024, marking a remarkable 142\% increase since 2020. However, streaming immersive videos is challenging as immersive videos are 4-6 times larger than conventional videos of the same perceived quality because of their panoramic nature~\cite{qian2018flare}. To tackle this issue, tile-based streaming~\cite{ye2023vrct}\cite{kan2022rart360}\cite{tu2023pstile} has emerged as an efficient solution to transmit immersive videos to reduce bandwidth consumption and improve Quality of Experience (QoE). In tile-based streaming, video chunks are spatially cropped into non-overlapping tiles, and only tiles inside the user’s predicted future viewports are prefetched at high bitrates, thus striking a good balance between bandwidth efficiency and user’s QoE.

The implementation of tile-based streaming requires two fundamental building blocks: \textit{viewport prediction} and \textit{bitrate selection}. In recent years, many neural adaptive methods have been proposed to design the two building blocks, which demonstrate superior performance over conventional methods by exploiting the strong non-linear fitting capability of neural networks (NNs). For instance, sequence-to-sequence models such as Gated Recurrent Unit (GRU)~\cite{guimard2022deep} and Long Short Term Memeory (LSTM)~\cite{rondon2021track}\cite{zhang2020epass360} have showcased higher accuracy for viewport prediction than traditional linear regression~\cite{hu2019vas360}. Additionally, deep reinforcement learning (DRL) algorithms~\cite{zhang2019drl360}\cite{jiang2020reinforcement}\cite{kan2022rart360} have demonstrated more potential for optimizing bitrate selection than heuristic or model-based algorithms as they can determine tile bitrates without any specific presumptions.

% Despite promising, existing learning-based methods have not fully captured the diversity in users' \textit{viewing patterns} and \textit{QoE requirements}, leading to \textit{poor generalization}. 
Despite promising, existing neural adaptive methods have not fully addressed the challenge of user diversity in \textit{viewing patterns} and \textit{QoE preferences}, leading to poor generalization.
On one hand, conventional wisdom trains the viewport prediction model with sets of users' viewport trajectories, but pays little attention to the prediction bias towards the training dataset~\cite{rondon2021track}\cite{zhang2019drl360}, which may cause significant accuracy loss. Specifically, users' viewing patterns naturally exhibit high diversity~\cite{jin2022you}\cite{lu2022personalized}. For example, some users may prefer to focus on particular objects, while others may prefer to explore the scene. In this context, when users' actual viewing patterns differ significantly from those observed in the training stage, the model may fail to predict accurately.
%, e.g., a user who prefers to focus on particular objects is predicted to view the entire scene. 
Such prediction bias limits the model's ability to serve users with diverse viewing patterns, resulting in poor generalization. %This therefore poses the \textbf{\textit{first challenge}}: \textit{how to reduce the prediction bias to improve the generalization of viewport prediction model?}

% On the other hand, conventional DRL-based algorithms regard a pre-defined QoE function as reward and train the bitrate selection model to optimize such function~\cite{zhang2019drl360}\cite{jiang2020reinforcement}. 
\Revision{On the other hand, conventional DRL-based algorithms define reward function as a QoE function with fixed weights to quantify the importance of different QoE metrics (e.g., video quality and rebuffering time)~\cite{zhang2019drl360}\cite{jiang2020reinforcement}. They then train the DRL model to optimize such function for bitrate selection. }
Nevertheless, different users often have different QoE preferences~\cite{li2023achieving}\cite{wang2019intelligent}. For instance, some users may prioritize high video quality, while others may prioritize smooth playback and tolerate quality distortion. 
% Therefore, it is difficult to characterize users’ diverse preferences with a fixed QoE function. 
\Revision{Therefore, it is difficult to characterize users' diverse QoE preferences with a fixed-weight QoE function, as the weights assigned to different QoE metrics can vary significantly among users with different preferences.}
As a result, the performance of existing DRL algorithms may significantly deteriorate when the optimized QoE function does not align to users’ actual preferences~\cite{li2022apprenticeship}. For instance, the DRL model trained to aggressively download low-bitrate tiles to avoid playback interruptions may select inappropriate bitrates for users that prioritize high video bitrates. Consequently, these DRL algorithms fail to generalize across users with diverse QoE preferences.

To tackle the challenge of user diversity, prior studies~\cite{zhang2019drl360}\cite{lu2022personalized}\cite{wang2017user} have attempted to categorize users into dinsinct groups based on their viewing and QoE preferences, and train personalized models for each group. 
% However, these approaches require retraining each time a new user group emerges, during which users will experience service degradation.
However, this approach necessitates the retraining of a new model whenever a new user group emerges, resulting in prohibitive training cost. 
As an alternative, researchers in~\cite{zhang2020epass360} have explored ensemble learning to improve the generalization of viewport prediction models, but their approach involves explicit model duplication, leading to substantial computational overhead.
For bitrate selection, recent works~\cite{li2022apprenticeship}\cite{wu2021paas} have proposed to train the DRL model with multiple QoE functions instead of a single one, but this approach suffers from catastrophic forgetting problem~\cite{yao2019adversarial} and may still experience performance degradation when serving users with QoE preferences unaligned to those optimized in the training stage.%but it still may suffer from performance degradation when serving users with QoE requirements unaligned to those in the training stage.

In this paper, we propose \texttt{MANSY}, an ense\textbf{M}ble and represent\textbf{A}tion lear\textbf{N}ing based \textbf{SY}stem for tile-based immersive video streaming, which addresses the user diversity challenge to improve generalization. 
To capture the viewing pattern diversity, we develop a viewport prediction model with an efficient Multi-viewport Trajectory Input Output (MTIO) architecture based on implicit ensemble learning (EL)~\cite{ganaie2022ensemble}\cite{havasi2021training}. 
% The MTIO architecture trains multiple sub-models with multiple input-output heads, with each head implicitly representing a sub-model. 
The MTIO architecture implicitly trains multiple sub-models with minor computation cost by establishing multiple input-output heads.
Each sub-model independently makes prediction, and their prediction results are ensembled to yield well-calibrated predicted viewports that reduce the prediction bias, thus leading to stronger generalization ability. 
Additionally, we also design the backbone of the prediction model based on Transformer~\cite{vaswani2017attention}, which leverages the attention mechanism to effectively learn long-term dependencies. This enables our model to predict the trends of viewport movements more accurately, further improving the prediction accuracy.

To accommodate users' diverse QoE preferences, we leverage the advanced representation learning (RepL) technique~\cite{bengio2013representation} to train the DRL bitrate selection model. Specifically, we encourage the model to mine useful hidden representations from users' QoE preferences, by incorporating mutual information into the reward function for training the model. 
In this way, we enable our model to capture essential characteristics of users' preferences, such as prioritization of bitrate quality and playback smoothness. This empowers our model to dynamically select bitrates based on users' QoE preferences, even when encountering those unseen during the training phase, thus achieving strong generalization.
Additionally, as directly computing mutual information is difficult, we further design an efficient NN model to estimate the mutual information term for reward calculation.

To summarize, this paper makes the following contributions:
% To summarize, the contributions of this paper are as follows:
\begin{itemize}
    \item We propose a novel tile-based immersive video streaming system \texttt{MANSY} that addresses the challenge of user diversity in both viewing patterns and QoE preferences to significantly improve generalization.

    \item We design an efficient MTIO-Transformer viewport prediction model based on implicit EL, which effectively reduces the prediction bias to serve users with various viewing patterns with minor computation cost.

    \item To the best of our knowledge, we are the first to combine RepL and DRL to train the bitrate selection model, enabling the model to maximize users' QoE with diverse preferences and thus achieving better generalization.
    \item Extensive experiments demonstrate the superiority of \texttt{MANSY} in both viewport prediction and bitrate selection. Results indicate that compared to state-of-the-art approaches, \texttt{MANSY} improves the mean prediction accuracy by \Revision{1.3\%-5.2\%/3.2\%-8.8\%} and mean QoE by 3.0\%-14.1\%/3.2\%-15.3\% on trained/unseen viewing patterns and QoE preferences, respectively. 

\end{itemize}

The rest of this paper is organized as follows. Section~\ref{sec:motivation} presents our observations on the impacts of user diversity, which motivates the design of our work. Section~\ref{sec:system_overview} provides the system overview of \texttt{MANSY}. Next, we elaborate the detailed design of the MTIO Tranformer viewport prediction model and RepL-enabled bitrate selection model in Section~\ref{sec:viewport_prediction} and~\ref{sec:bitrate_selection}, respectively. We then conduct extensive experiments to evaluate the performance of \texttt{MANSY} in Section~\ref{sec:evaluation}. The related work is provided in Section~\ref{sec:related_work}. \Revision{We also discuss some potential methodologies to further improve \texttt{MANSY} in Section~\ref{sec:discussion}.} Finally, Section~\ref{sec:conclusion} concludes this paper.

The codes associated with this article are publicly available at \url{https://github.com/duowuyms/MANSY_ImmersiveVideoStreaming}.

\section{Motivation and Analysis}
\label{sec:motivation}
\subsection{Impact of Viewing Pattern Diversity}
The panoramic nature of immersive videos allows users to freely rotate their heads to watch the most attractive parts of the videos. As different users often have different viewing preferences, users' viewing patterns naturally exhibit high diversity~\cite{jin2022you}\cite{lu2022personalized}, posing unique challenge to the design of viewport prediction model.

\begin{figure}[t]
    \centering
    \begin{minipage}{0.2\textwidth}
    \centering
    \subfigure{
    \includegraphics[width=\textwidth]{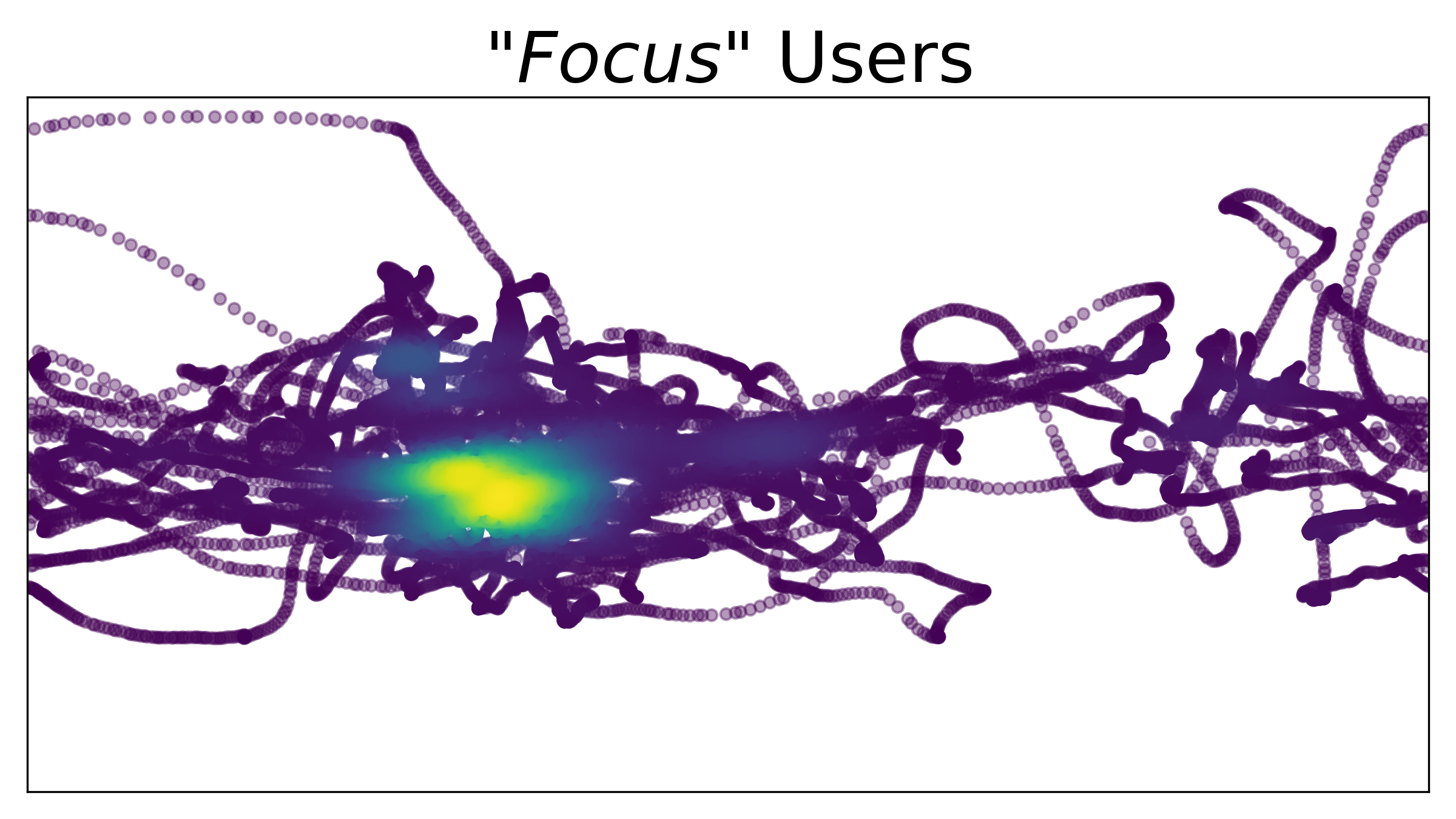}
    \label{subfig:moti_vp_group1}
    }
    \centering
    \subfigure{
    \includegraphics[width=\textwidth]{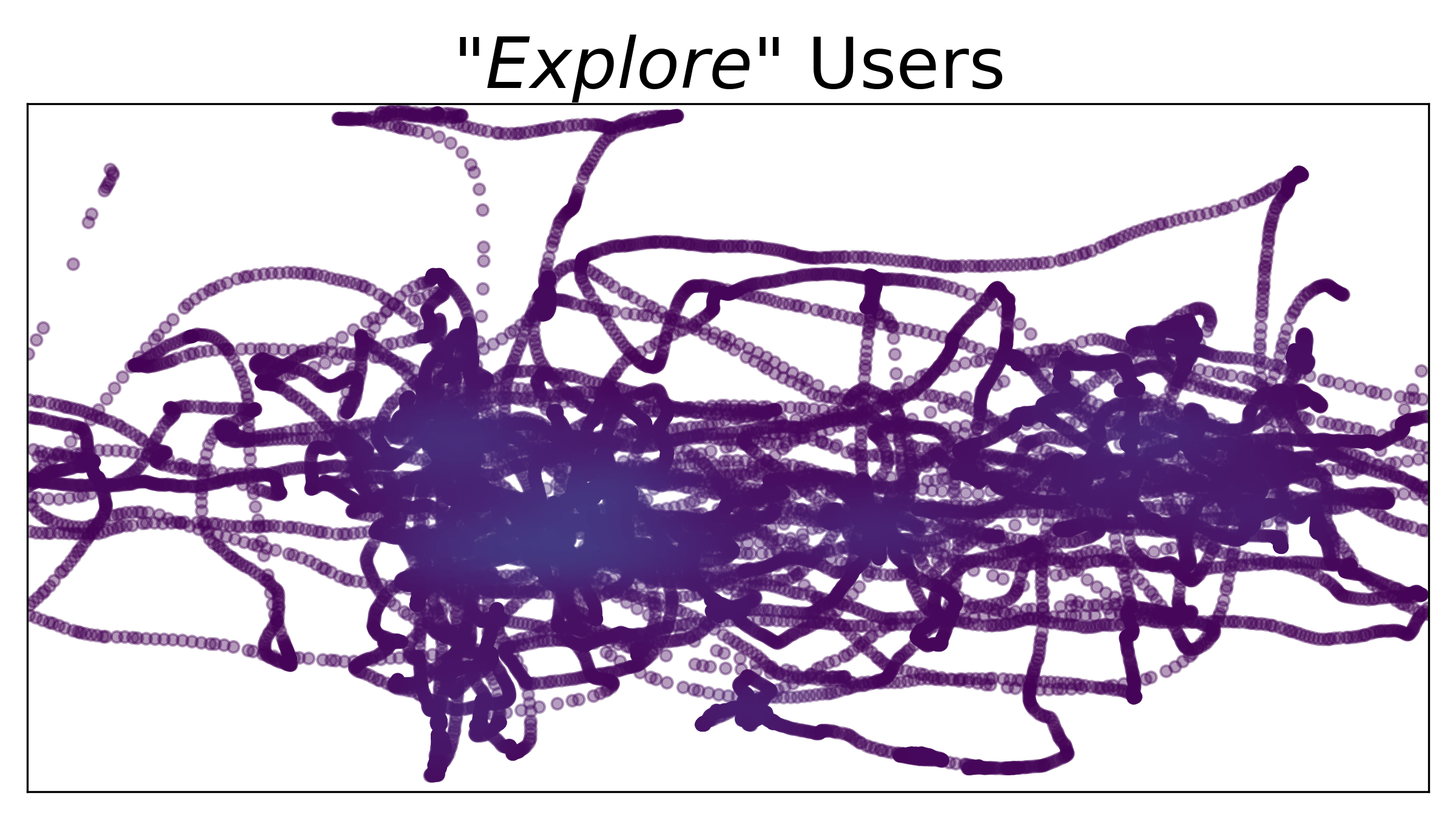}
    \label{subfig:moti_vp_group2}
    }
    \end{minipage}
    \begin{minipage}{0.2\textwidth}
    \centering
    \subfigure{
    \includegraphics[width=\textwidth]{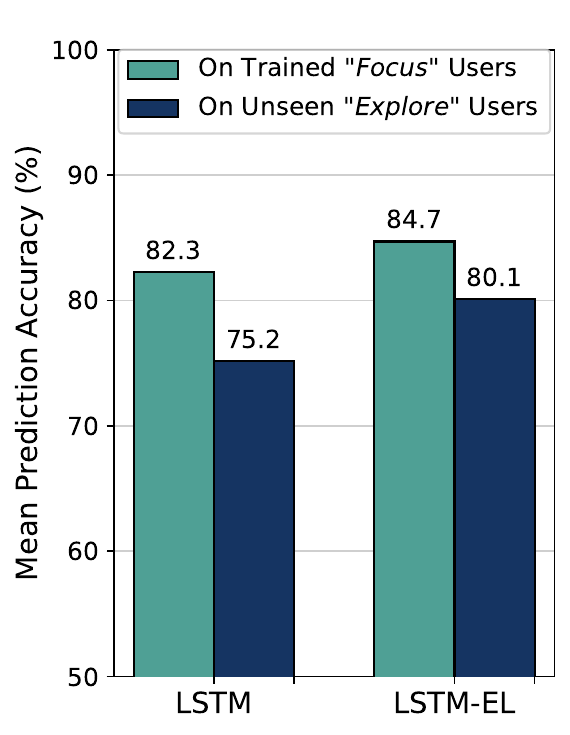}
    \label{subfig:moti_vp_acc}
    }
    \end{minipage}
    \vspace{-0.3cm}
    % \caption{Illustration of the impact of viewing pattern diversity and the effectiveness of ensemble learning (EL). \textit{Left figure}: the density maps of two type of users' viewport directions. \textit{Right figure}: comparison of prediction performance with/without EL.}
    \caption{Viewport prediction accuracy on two set of users with different viewing patterns.}
    \label{fig:moti_vp}
    \vspace{-0.3cm}
\end{figure}

Conventional approaches~\cite{rondon2021track}\cite{zhang2019drl360} simply trains time-series NN models for viewport prediction with a set of collected users' viewport samples, but neglect the model prediction bias towards the training dataset. When users' actual viewing patterns differ significantly from those observed in the training dataset, their models may suffer from significant accuracy loss, resulting in poor generalization.
We implement an LSTM model~\cite{zhang2019drl360} to demonstrate the ineffectiveness of these approaches. 
We set the historical and prediction window to be 1 second, and use 8x8 as the tiling scheme. We consider prediction accuracy as the performance metric, which is calculated as the intersection of union of the predicted and ground-truth viewports~\cite{rondon2021track}. Besides, two set of users' viewports are sampled from a open dataset~\cite{wu2017dataset}: the \textit{Focus} users prefer to focus on the objects in the video center, while the \textit{Explore} users prefer to explore the entire scene, as depicted in Figure~\ref{fig:moti_vp}. 
We use the \textit{Focus} set for training and consider the \textit{Explore} set as users with unseen viewing patterns.
The measurement results are presented in Figure~\ref{fig:moti_vp}. As depicted, \textit{LSTM} achieves a high prediction accuracy of 82.3\% on the trained \textit{Focus} users. However, when confronted with users exhibiting significantly different viewing patterns, it struggles to accurately predict their viewport movements due to the inherent bias towards the training dataset. This leads to a noticeable performance drop on unseen \textit{Explore} users, with an absolute accuracy loss up to 7.2\%.

One way to combat the above limitation is to train the model over a large diverse dataset, but this requires tremendous amount of data with rich statistical diversity~\cite{lu2022personalized}, which, however, is practically unavailable. 
Grouping users based on their viewing patterns and training personalized models for each group seems plausible~\cite{lu2022personalized}\cite{wang2017user}, but this necessitates retraining of a new model whenever a new group emerges, leading to prohibitive training cost. %This leads to prohibitive training cost and potential poor viewing experiences for users before the retraining is complete.%, and users may suffer from poor viewing experience due to inaccurate viewport prediction before the retraining completes. 
Previous work~\cite{zhang2020epass360} has showcased the potential of ensemble learning (EL) to reduce prediction bias and improve model generalization. It works by training multiple independent sub-models and combining their predictions to yield more accurate results. 
% While conceptually easy, the success of EL to improve generalization has been supported by numerous works and theories~\cite{ganaie2022ensemble}. 
The rationale behind is that when confronted with unknown data samples, the ensemble members may exhibit bias towards different directions, but their ensembled predictions can yield well-calibrated results that offset the bias~\cite{ganaie2022ensemble}\cite{turkoglu2022film}. For illustrative purpose, we also implement an ensembled-version of LSTM model (denoted as \textit{LSTM-EL}) following the approach in~\cite{zhang2020epass360}, which includes three independent LSTM sub-models. As shown in Figure~\ref{fig:moti_vp}, \textit{LSTM-EL} is more robust to unseen viewing patterns: it maintains a high prediction accuracy of 80.2\% on unseen \textit{Explore} users with a smaller accuracy drop (4.2\%) compared to \textit{LSTM} (7.2\%).

Despite promising, the challenge of EL is the substantial computation overhead caused by explicit model duplication, hindering its deployment on resource-constrained client devices. For instance,  compared with \textit{LSTM}, \textit{LSTM-EL} introduce an extra overhead of 200\% in terms of model parameters and floating-point operations. To tackle this challenge, we design an efficient Multi-viewport Trajectory Input Output (MTIO) architecture based on implicit EL. The MTIO architecture implicitly trains multiple sub-models without model duplication by establishing multiple input-output heads, thus exploiting the benefit of EL with negligible overhead. Moreover, we design the model architecture based on Transformer, which further improves the predictive performance of our model.

\begin{figure}[t]
    \centering
     \includegraphics[width=0.4\textwidth]{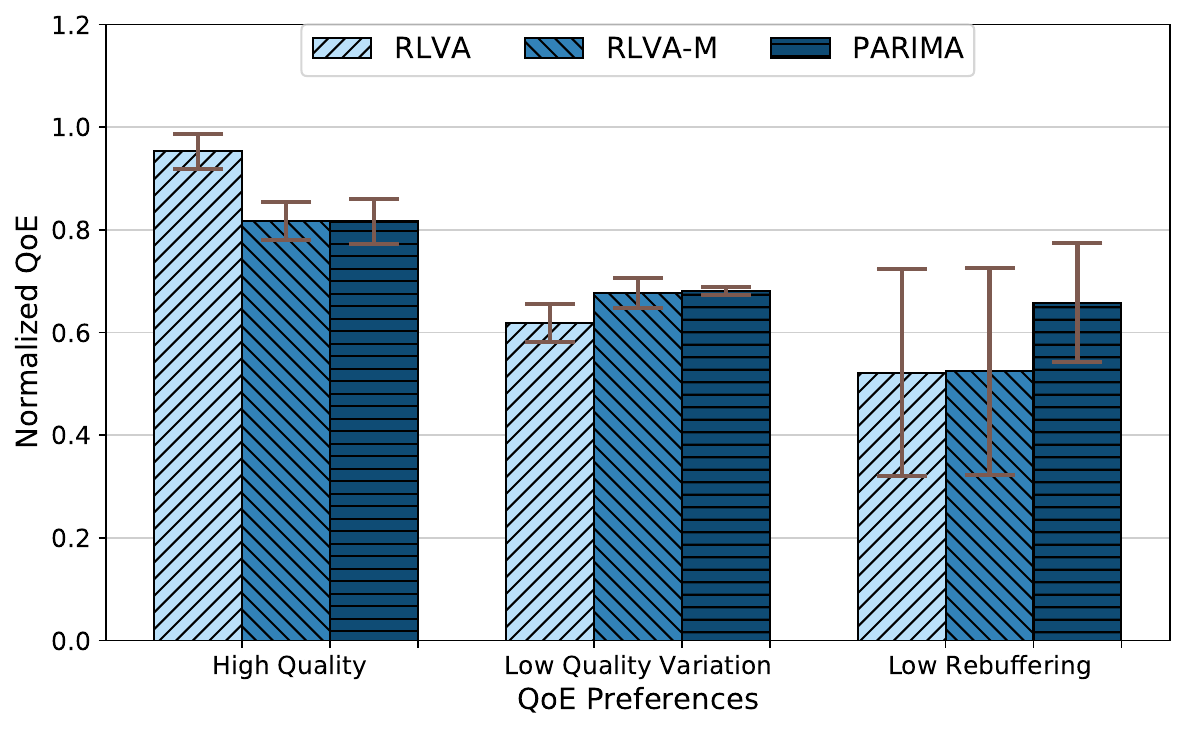}
    \vspace{-0.2cm}
    % \caption{Illustration of the limitation of existing DRL-based solutions in handling QoE preference diversity. \textit{RLVA} is the DRL method trained on \textit{High Quality} preference; \textit{RLVA-M} is trained on both \textit{High Quality} and \textit{Low Quality Variation}; \textit{PARIMA} is a heuristic method.}
    \caption{Performance of bitrate selection methods on three different QoE preferences.}
    \label{fig:moti_abr_performance}
    \vspace{-0.2cm}
\end{figure}

\subsection{Impacts of QoE Preference Diversity}
\label{subsec:motivation_abr}

During the streaming of immersive videos, users' QoE is characterized by various metrics, such as video quality, rebuffering time, and quality variation~\cite{chen2023live360}\cite{jin2023ebublio}. To effectively optimize multiple metrics, one common practice is to define a linear function that assigns different weights to each metric. Since users often have different QoE preferences (e.g., prioritizing high quality or low rebuffering), the assigned weights can vary among users~\cite{li2023achieving}\cite{li2022apprenticeship}. However, previous DRL methods~\cite{zhang2019drl360}\cite{jiang2020reinforcement}\cite{kan2022rart360} consider a fixed-weight QoE function as the reward to train the bitrate selection model, which limits their ability to generalize across diverse QoE preferences. Consequently, the discrepancy between users' actual QoE preferences and the optimized one can significantly degrade their performance~\cite{li2022apprenticeship}, resulting in a poor video watching experience for users. Existing works~\cite{li2022apprenticeship}\cite{wu2021paas} have proposed training the DRL model with multiple QoE functions simultaneously to address this challenge. However, this naive approach suffers from the catastrophic forgetting problem~\cite{yao2019adversarial}, where the knowledge learned to optimize previous QoE preferences is overwritten by knowledge learned for current ones. Moreover, it may still experience performance degradation when serving users with unseen QoE preferences that differ from those optimized during training.

To illustrate the limitation of existing DRL-based solutions, we take \textit{RLVA}~\cite{kan2022rart360}, a state-of-the-art DRL bitrate selection model that optimizes a single QoE preference, as an example. We train \textit{RLVA} on the \textit{High Quality} QoE preference and regard \textit{Low Quality Variation} and \textit{Low Rebuffering} as unseen preferences (detailed descriptions of QoE preferences are provided in Section 5). For comparative analysis, we also implement a heuristic approach \textit{PARIMA}~\cite{chopra2021parima} and report its performance across the three preferences, as shown in Figure~\ref{fig:moti_abr_performance}.

\begin{figure*}[t]
    \centering
    \includegraphics[width=0.95\textwidth]{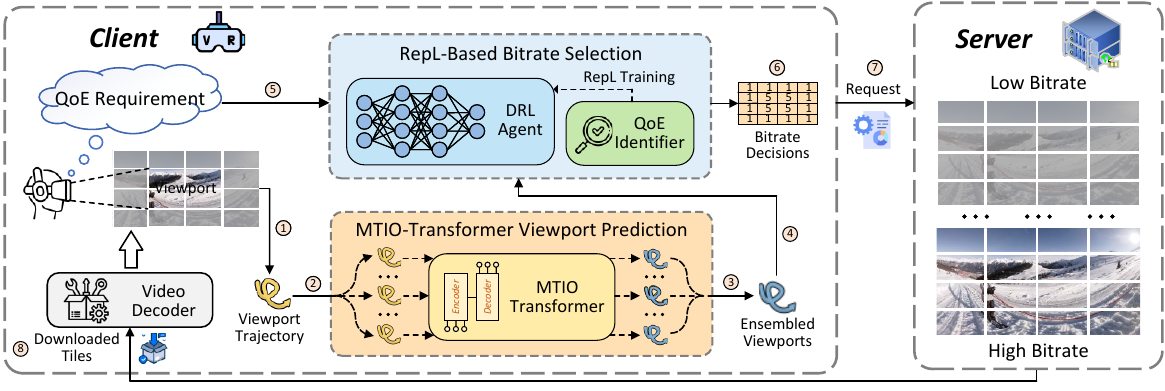}
    \vspace{-0.2cm}
    \caption{System framework of the proposed tile-based immersive video streaming system \texttt{MANSY}.}
    \label{fig:system_framework}
    \vspace{-0.2cm}
\end{figure*}

As expected, \textit{RLVA} exhibits superior performance to \textit{PARIMA} on the trained \textit{High Quality} preference, with an average QoE increase of 16.7\%. However, its performance significantly deteriorates on the unseen preferences, resulting in lower QoE compared to \textit{PARIMA}. The poor performance of \textit{RLVA} is attributed to its utilization of a fixed QoE function, which fails to guide the model in selecting appropriate bitrates for different preferences. Next, we further train \textit{RLVA} using both \textit{High Quality} and \textit{Low Quality Variation} preferences (referred to as \textit{RLVA-M}) to demonstrate the ineffectiveness of naive training with multiple QoE functions. As depicted in Figure~\ref{fig:moti_abr_performance}, while \textit{RLVA-M} achieves comparable performance to \textit{PARIMA} on the trained \textit{Low Quality Variation} preference, it experiences performance degradation on the trained \textit{High Quality} preference compared to \textit{RLVA}. This degradation stems from the forgetting problem, as the knowledge acquired to optimize \textit{High Quality} is overwritten by the knowledge to optimize \textit{Low Quality Variation}. Additionally, when serving users with the unseen \textit{Low Rebuffering} preference, \textit{RLVA-M} still performs worse than \textit{PARIMA}, achieving poor generalization.

The key to optimize diverse QoE preferences is to learn useful representations about the relationship between QoE and the selected bitrates. Hence, in this paper, we leverage the advanced representation laerning (RepL)~\cite{bengio2013representation} technique to tackle the QoE preference diversity challenge. Specifically, we augment the reward function for model training with mutual information, which encourages the model to learn hidden representations that expose salient attributes of users' QoE preferences. The learned useful representations empowers our model to dynamically select bitrates based on users' preferences and generalize to diverse preferences including those unseen during the training stage.

\section{System Overview}
\label{sec:system_overview}
Figure~\ref{fig:system_framework} depicts the system overview of \texttt{MANSY}, which comprises two core components: MTIO-Transformer viewport prediction and RepL-based bitrate selection.

\textbf{MTIO-Transformer viewport prediction.} On the client side, a user watches an immersive video with a head-mounted device (HMD) such as Facebook Oculus and Microsoft Hololens, which continuously records the user's viewport trajectory. 
The historical viewport trajectory is extracted from the HMD and passed to the multiple input heads of the MTIO-Transformer. Based on the received trajectory, the model outputs multiple trajectories and ensembles them to generate an accurate prediction with small bias. The ensembled trajectory is subsequently passed to the bitrate selection module as an important reference for determining tile bitrates.
% All the outputs are then ensembled to generate accurate and unbiased predicted viewports, which are subsequently used as an important reference for determining tile bitrates in the bitrate selection module.

\textbf{RepL-based bitrate selection.} Given the predicted viewports as well as user's QoE preference information, a DRL agent parameterized by an NN model is used to dynamically select tile bitrates based on the environment state (e.g., network bandwidth conditions and playback buffer size). 
Once the agent makes the bitrate decisions, the HMD then requests tiles at the corresponding bitrates, downloads tiles from the server and displays them to the user. 
Note that the agent is offline trained with the RepL technique, where a QoE identifier is designed to facilitate the agent to disentangle highly semantic and useful representations from the QoE preference information. The QoE identifier is essentially another NN model used to guide the agent to maximize the mutual information between agent's selected bitrates and user's QoE preference. 

\section{MTIO-Transformer Viewport Prediction}
\label{sec:viewport_prediction}

Figure~\ref{fig:mimo_lstm} depicts the architecture of the viewport prediction model of \texttt{MANSY}, which incorporates two core designs:

\begin{itemize}
    \item  \textit{MTIO architecture.} We design the model with a Multi-viewport Trajectory Input Output (MTIO) architecture to efficiently reduce the prediction bias, so that our model can generalize across a broad range of users with diverse viewing patterns. Our model utilizes the insight that a neural network is over-parameterized and has sufficient capacity to fit multiple sub-networks simultaneously~\cite{havasi2021training}. 
    % It therefore trains multiple independent sub-models with multiple input-output heads, with each head implicitly representing a sub-model. 
    It therefore trains multiple independent sub-models within one network by establishing multiple input-output heads, with each head implicitly representing a sub-model.
    While each sub-model (i.e., head) may exhibit bias, the ensemble of their predictions can result in well-calibrated outcomes that effectively reduce such bias~\cite{ganaie2022ensemble}\cite{turkoglu2022film}, thus leading to improved generalization. 
    Moreover, since sub-models are trained without explicit duplication, our model can utilize the benefits of ensemble with negligible overhead.
    \item \textit{Transformer-based backbone.} 
    % We further design the backbone of our model based on Transformer. 
    % It leverages the attention mechanism~\cite{vaswani2017attention} to effectively learn long-term dependencies and mine users' viewing patterns, thus further improving the predictive performance.
    Considering that long-term dependencies greatly influence time series prediction tasks including viewport prediction, we further design the backbone of our model based on Transformer. It leverages the attention mechanism~\cite{vaswani2017attention} to effectively learn long-term dependencies to predict the trends of viewport movements more accurately, thus further improving the predictive performance.
    % The heart of the backbone is the attention mechanism~\cite{vaswani2017attention}, which enables our model to effectively learn long-term dependencies information and mine users' viewing patterns, further improving the predictive performance.
\end{itemize}

\subsection{Model Design}
Let $\bold{v}_t=(x_t, y_t)$ denote the user's viewport at timestep $t$, where $x_t, y_t$ represent the horizontal and vertical coordinates of viewport center in the equirectangular projection of the video\footnote{The concepts in this section can be easily extended to other forms of coordinates, such as Euler angles and quaternion.}, respectively. 
% Define $M$ as the number of input-output heads. 
The detailed design of the model is explained as follows.

\begin{figure}[t]
    \centering
    \includegraphics[width=0.46\textwidth]{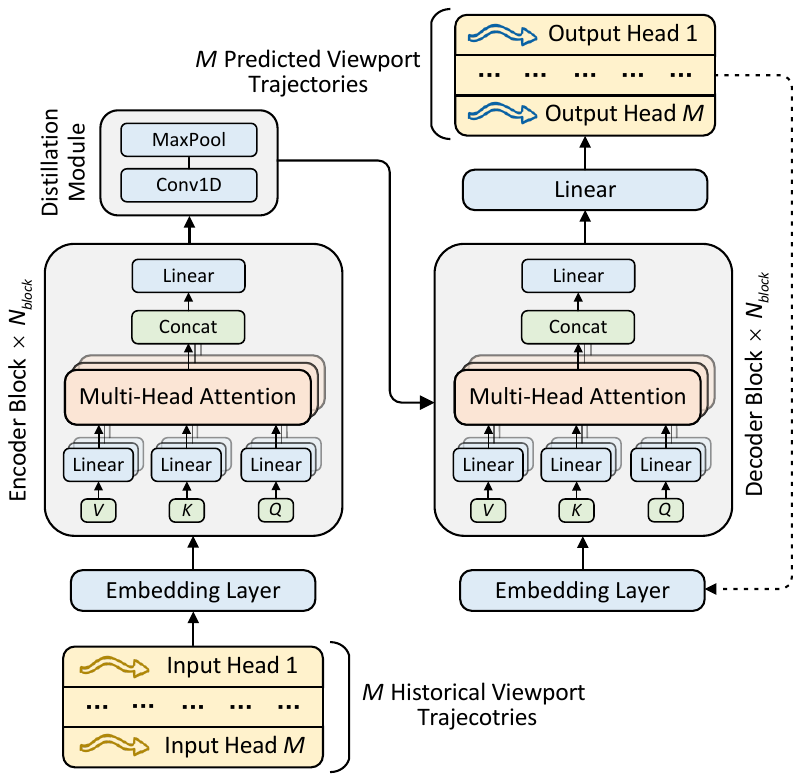}
    \caption{The architecture of the proposed MTIO-Transformer viewport prediction model.}
    \label{fig:mimo_lstm}
    \vspace{-0.5cm}
\end{figure}

\textbf{Multi-head inputs.} As shown in Figure~\ref{fig:mimo_lstm}, our model incorporates $M$ input heads. At any timestep $t$ during video playback, it takes $M$ historical viewport trajectories $\{\bold{\hat{v}}^i_{t-h}, \cdots, \bold{\hat{v}}^i_t\}_{i=1}^M$ as inputs, where $h$ represents the historical time horizon. To ensure each head is independently trained, during the training stage, the parameters of each head are randomly initialized and the $M$ historical trajectories are randomly sampled from the training dataset. All trajectories are stacked, projected into a sequence of $d_e$-dimension embeddings and passed to the backbone network to extract underlying features.

\textbf{Encoder-decoder.} 
% The input embeddings are fed to the encoder with $N_{block}$ stacked blocks to extract complex features. 
The backbone network adopts an encoder-decoder architecture, as illustrated in Figure~\ref{fig:mimo_lstm}. Both the encoder and decoder consist of $N_{block}$ stacked blocks to extract complex features.
The core of each block is the attention mechanism which enables effectively learning of long-term dependency information from the input embeddings. The attention mechanism can be described as mapping queries and sets of key-value pairs to attention weights that are assigned to different elements of the input sequence~\cite{vaswani2017attention}. Specifically, let  $Q\in \mathbb{R}^{d_k\times d_e}, K\in\mathbb{R}^{d_k\times d_e}, V\in\mathbb{R}^{d_v\times d_e}$ represent the query, key and value matrix, respectively. The attention weights are computed by:
\begin{equation}
    \setlength{\abovedisplayskip}{3pt}
    % \text{Attention}(Q,K,V) = \text{softmax}({QK^T}/\sqrt{d_k})V
    Attention(Q,K,V) = \text{softmax}({QK^T}/\sqrt{d_k})V
    \setlength{\belowdisplayskip}{3pt}
\end{equation}
where $\sqrt{d_k}$ is the scale factor. The attention mechanism can be repeated by $N_{ah}$ heads with different subspaces of $Q, K,V$, which benefits the model to jointly consider information from different representation subspaces. The multi-head attention is achieved by:
\begin{equation}
    \begin{aligned}
    \setlength{\abovedisplayskip}{3pt}
    % \text{MultiHead}(Q, K,V) = \text{concat}([\text{attn\_head}_j]^{N_{ah}}_{j=1})W^O, \\
    % \text{attn\_head}_j = \text{Attention}(QW^Q_j, KW^K_j, VW^V_j)
    MultiHead(Q, K,V) = \text{concat}([attn\_head_j]^{N_{ah}}_{j=1})W^O, \\
    attn\_head_j = Attention(QW^Q_j, KW^K_j, VW^V_j)
    \setlength{\belowdisplayskip}{3pt}
    \end{aligned}
\end{equation}
where $W_j^Q \in \mathbb{R}^{d_e\times d_k}, W_j^K \in \mathbb{R}^{d_e\times d_k}, W_j^V \in \mathbb{R}^{d_e\times d_v}$ and $W^O\in \mathbb{R}^{N_{ah}d_v\times d_e}$ are all learnable weight matrices.

The encoder generates a sequence of hidden features extracted from the historical viewports, which, however, may contain redundant information~\cite{zhou2021informer}. To address this issue, rather than directly feeding the entire sequence of features to the decoder, we further design a distillation module to prioritize the dominant features from the encoder outputs. The distillation module comprises a 1D convolution layer and a max-pooling layer, as shown in Figure~\ref{fig:mimo_lstm}. It is used to compress the sequence length of the encoder outputs, resulting in a more focused set of input features for the decoder. This enhancement enables the decoder to better capture the viewport moving patterns. Additionally, another benefit of the distillation process is the reduction in computational workload for the decoder, as the smaller input feature length requires less computation.

\textbf{Multi-head outputs.} 
The model generates the future predicted viewports by linearly projecting the outputs of the decoder. In particular, our model is designed with $M$ output heads. Each output head produces a viewport trajectory that corresponds to the prediction result of the corresponding input head. In addition, our model predicts $M$ viewport trajectories in the autoregressive manner. It progressively predicts the viewports of next timestep by repeatedly injecting the previous predictions as inputs of the decoder. As a result, the future $M$ viewport trajectories $\{\bold{v}^i_{t+1}, \cdots, \bold{v}^i_{t+H}\}_{i=1}^M$ are produced by the model, where $H$ denotes the prediction horizon.

\begin{figure*}[t]
    \centering
    \includegraphics[width=0.95\textwidth]{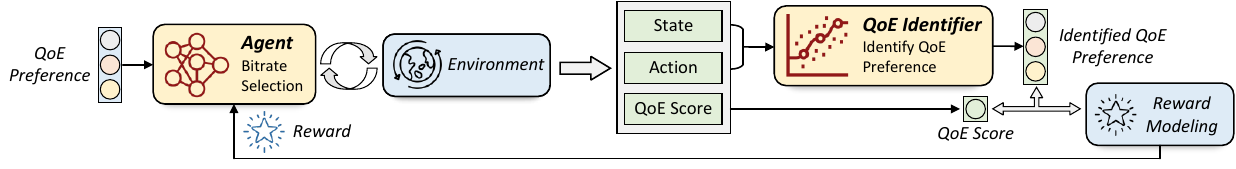}
    \caption{Illustrations of the proposed RepL-based learning framework. Note that the reward for training the agent is composed of two parts, which are derived from the QoE score and outputs of QoE identifier.}
    \vspace{-0.5cm}
    \label{fig:training_framework}
\end{figure*}

\textbf{Loss function.} The model is trained to minimize the distance between its predicted viewports $\{\bold{v}^i_{t+1}, \cdots, \bold{v}^i_{t+H}\}_{i=1}^M$ and ground truth viewports $\{\bold{\hat{v}}^i_{t+1}, \cdots, \bold{\hat{v}}^i_{t+H}\}_{i=1}^M$. 
% To accurately measure the distance between two viewports, we design a distance function based on mean square error, which considers the periodicity of viewport coordinates. 
To accurately measure the distance between two viewports, we design a distance function based on mean square error, which considers the periodicity of viewport horizontal coordinate~\cite{chopra2021parima}.  
Specifically, let $\bold{v}=({x}, y), \bold{\hat{v}}=(\hat{x},\hat{y})$ represent the predicted and ground truth viewports, respectively, their distance is calculated by:
\begin{equation}
    \begin{aligned}
    \setlength{\abovedisplayskip}{3pt}
    % \text{Dist}(\bold{\hat{v}, v}) &= \big(\text{Dist}_x(\hat{x},x) ^2+\text{Dist}_y(\hat{y},y)^2\big) /2 \\
    % \text{Dist}_x(\hat{x},x) &= \min(|\hat{x}-x|, |\hat{x}+w-x|,|\hat{x}-w-x|) \\
    % \text{Dist}_y(\hat{y},y) &= \min(|\hat{y}-y|, |\hat{y}+h-y|,|\hat{y}-h-y|)
    Dist(\bold{v, \hat{v}}) &= \big(Dist_x(x,\hat{x}) ^2+Dist_y(y,\hat{y})^2\big) /2 \\
    Dist_x(x,\hat{x}) &= \min(|x-\hat{x}|, |x+w-\hat{x}|,|x-w-\hat{x}|) \\
    % Dist_y(y,\hat{y}) &= \min(|y-\hat{y}|, |y+h-\hat{y}|,|y-h-\hat{y}|)
    Dist_y(y,\hat{y}) &= |y-\hat{y}|
    \setlength{\belowdisplayskip}{3pt}
    \end{aligned}
\end{equation}
% where $w$ and $h$ are the width and height of the video, respectively. 
where $w$ are the width and height of the video, respectively. 
The loss function is then defined as the sum of distances of each input-output head:
\begin{equation}
    \setlength{\abovedisplayskip}{3pt}
    % \text{Loss} = \sum_{i=1}^M\text{Dist}(\bold{\hat{v}^i, v^i})
    Loss = \sum_{i=1}^M Dist(\bold{v^i, \hat{v}^i})
    \setlength{\belowdisplayskip}{3pt}
\end{equation}

\subsection{Ensembling Predictions}
During the inference phase, the model extracts a historical viewport trajectory $\{\bold{\hat{v}}'_{t-h}, \cdots, \bold{\hat{v}}'_t\}$ from the HMD to predict the future viewports. This trajectory is duplicated by $M$ times and fed to each input head, i.e., $\{\bold{\hat{v}}^i_{t-h}, \cdots, \bold{\hat{v}}^i_t\}_{i=1}^M=\{\bold{\hat{v}}'_{t-h}, \cdots, \bold{\hat{v}}'_t\}$. The output heads will produce $M$ independent prediction results for the same input. Their predictions are finally ensembled to yield calibrated viewports that effectively reduce the prediction bias to improve the predictive performance:
\begin{equation}
    \setlength{\abovedisplayskip}{3pt}
    % \bold{\hat{v}}'^e_{t+j}=\frac{1}{M}\sum_{i=1}^M \bold{\hat{v}}'^i_{t+j}, \forall j\in\{1,\cdots,H\}
    \bold{v}'^e_{t+j}=\frac{1}{M}\sum_{i=1}^M \bold{v}'^i_{t+j}, \forall j\in\{1,\cdots,H\}
    \setlength{\belowdisplayskip}{3pt}
\end{equation}
where $\bold{v}'^e_{t+j}$ represents the ensembled predicted viewport. 

Note that having $M$ input-output heads only introduces negligible computation overhead. 
% For example, in the later experiment settings, we find that the proposed MTIO architecture only introduces $1.5\%$ increase in the total number of model parameters and $0.8\%$ increase in floating-point operations (FLOPs). 
This is because the MTIO architecture only requires additional model parameters in the input and output layers, and meanwhile the model can obtain well-calibrated predictions in just a single forward pass. The detailed analysis of computation overhead of MTIO is covered in Section~\ref{subsec:evaluate_viewport_prediction}.

\section{RepL-based Bitrate Selection}
\label{sec:bitrate_selection}

In this section, we present the detailed design of the bitrate selection approach of \texttt{MANSY}. Figure~\ref{fig:training_framework} illustrates the proposed RepL-based learning framework, which compromises a DRL agent for bitrate selection and a QoE identifier model to facilitate the agent to learn useful representations of users' QoE preferences. Under the current state of the environment, the agent generates a bitrate action and receives a QoE score as a partial reward signal. Notably, the outputs of the QoE identifier also serves as a reward signal, which captures the mutual information between users' QoE preferences and agent's selected bitrates. The final reward to train the agent is then derived from the combination of QoE score and outputs of QoE identifier.

In the following, we begin by describing the QoE model that characterizes the users' QoE preferences, then we elaborate the details of the proposed RepL-based bitrate selection algorithm, including DRL agent design, mutual-information-based reward modeling, as well as training methodology.

\subsection{Quality of Experience Model}
Following previous works~\cite{jiang2020reinforcement}\cite{lu2022personalized}, we adopt three critical metrics to characterize the user's QoE, namely average viewport quality, quality variation and rebuffering time. These metrics are defined as follows:

\textbf{1) Average viewport quality.} The average viewport quality $QoE^1_c$ describes the average bitrate quality of tiles inside user's actual viewport of chunk $c$. 
% Assume that $N_{tile}$ is the number of tiles which are indexed in raster-scan order and $\mathcal{R}$ denotes the set of bitrate versions to encode the tiles. Let $\hat{v}_{c,i}$ represent whether $i$-th tile of chunk $c$ is inside user's actual viewport and $r_{c,i}\in\mathcal{R}$ stand for the bitrate allocated to the tile. 
This metric can be calculated by:
\begin{equation}
    \setlength{\abovedisplayskip}{3pt}
    % \text{Loss} = \sum_{i=1}^M\text{Dist}(\bold{\hat{v}^i, v^i})
    QoE^1_c=\frac{\sum_{i=1}^{N_{tile}}\hat{v}_{c,i}r_{c,i}}{\sum_{i=1}^{N_{tile}}\hat{v}_{c,i}}
    \setlength{\belowdisplayskip}{3pt}
\end{equation}
where $N_{tile}$ is the total number of tiles; $\hat{v}_{c,i}$ is a boolean variable indicating whether $i$-th tile of chunk $c$ is inside user's actual viewport; $r_{c,i}$ stands for the bitrate allocated to $i$-th tile.
% Note that $\hat{v}_{c,i}$ is a binary variable: if $i$-th tile is inside user's actual viewport, $\hat{v}_{c,i}=1$; otherwise, $\hat{v}_{c,i}=0$.

\textbf{2) Quality variation.} The variations of tile quality inside the viewport and the viewport quality between two consecutive chunks should be smooth to avoid causing sickness or headache to users. The viewport quality variation $QoE^2_c$ is measured by:
\begin{equation}
    \setlength{\abovedisplayskip}{3pt}
    % \text{Loss} = \sum_{i=1}^M\text{Dist}(\bold{\hat{v}^i, v^i})
    QoE^2_c=\frac{\sum_{i=1}^{N_{tile}}|\hat{v}_{c,i}r_{c,i} - QoE^1_c|}{\sum_{i=1}^{N_{tile}}\hat{v}_{c,i}} + |QoE^1_c-QoE^1_{c-1}|
    \setlength{\belowdisplayskip}{3pt}
\end{equation}
where the first term denotes the intra-variation of tile quality inside the viewport, and the second term denotes the inter-variation of viewport quality between consecutive chunks.

\textbf{3) Rebuffering time.} If the playback buffer goes empty before a chunk is downloaded, the user will suffer from a \textit{rebuffering} event. 
Following previous methods~\cite{zhang2019drl360}\cite{lu2022personalized}, we calculate the rebuffering time $QoE^3_c$ by:
\begin{equation}
    \setlength{\abovedisplayskip}{3pt}
    % \text{Loss} = \sum_{i=1}^M\text{Dist}(\bold{\hat{v}^i, v^i})
    QoE^3_c=(l_c-b_c)_+
    \setlength{\belowdisplayskip}{3pt}
\end{equation}
where $l_c$ is the download time of chunk $c$; $b_c$ denotes the buffer occupancy when the download request of chunk $c$ is sent; $(\,\cdot\,)_+$ represents the function of $\max(\,\cdot\,, 0)$.

Based on the above metrics, user's QoE for $c$-th chunk can be modeled as:
\begin{equation}
    \setlength{\abovedisplayskip}{3pt}
    % \text{Loss} = \sum_{i=1}^M\text{Dist}(\bold{\hat{v}^i, v^i})
    QoE_c=\lambda_1 QoE^1_c-\lambda_2QoE^2_c-\lambda_3QoE^3_c
    \setlength{\belowdisplayskip}{3pt}
\end{equation}
Here, $\lambda_1, \lambda_2, \lambda_3$ are non-negative weight parameters that measure the relative importance of different metrics and satisfy $\lambda_1+\lambda_2+\lambda_3=1$. Therefore, we use $w=(\lambda_1,\lambda_2,\lambda_3)$ to represent the user's QoE preference. Similar models have also been adopted in~\cite{jiang2020reinforcement}\cite{lu2022personalized}\cite{wu2021paas}.

\subsection{DRL Agent Design}
The DRL agent is responsible for bitrate allocation based on the user's QoE preference and environment state information. The inputs, outputs, and NN architecture of the agent are designed as follows.

\textbf{Inputs.} When determining tile bitrates for chunk $c$, the agent takes a QoE preference $w$ and information of environment state $s_c$ as inputs. Formally, the environment state $s_c$ is modeled as:
\begin{equation}
    \setlength{\abovedisplayskip}{3pt}
    {s}_c=({Z}_c, {R}_c, \vec{{v}}_c, \vec{g}_c, \vec{n}_c, \vec{q}^{\,1}_c, \vec{q}^{\,2}_c, \vec{q}^{\,3}_c, b_c) \notag
    \setlength{\belowdisplayskip}{3pt}
\end{equation}
Here, ${Z}_c$ and ${R}_c$ record the sizes and bitrate qualities of each tile at different bitrate versions, respectively. $\vec{{v}}_c$ is the binary vector that indicates whether a tile is inside the predicted viewport. $\vec{g}_c$ denotes the viewport prediction accuracy of past $k$ chunks. $\vec{n}_c$ is the vector of past $k$ measured network throughputs. $\vec{q}^{\,1}_c, \vec{q}^{\,2}_c, \vec{q}^{\,3}_c$ record the average viewport quaity, quality variation and rebuffering time of the past $k$ chunks, respectively. Finally, $b_c$ represents the buffer occupancy.

\textbf{Outputs.} Based on QoE preference $w$ and state $s_c$, the agent outputs an action $a_c$ that corresponds to the bitrates allocated to tiles inside and outside the predictied viewport. The action $a_c$ is represented as $a_c=(r_c^{in}, r_c^{out})$, where $r_c^{in}, r_c^{out}\in \mathcal{R}$ and $\mathcal{R}$ denotes the discrete candidate bitrate set of tiles. 
In particular, $r_c^{in}, r_c^{out}$ satisfy the constraint of $r_c^{in} \ge r_c^{out}$, as bitrate of tiles inside the viewport should be larger than that outside the viewport.
All possible combinations of $r_c^{in}, r_c^{out}$ constitute the discrete action space. 
Based on the output action, we employ a pyramid-based strategy to assign bitrates to tiles according to their distance to the predicted viewport. Specifically, we assign $r^{in}_c$ to tiles inside the predicted viewport, then iteratively scale the boarder of viewport with one tile, assign $r^{out}_c/scale$ to the newly covered tiles\footnote{If $r_c^{out} / scale\notin \mathcal{R}$, we replace the $r_c^{out} / scale$ with the closest bitrate version in $\mathcal{R}$ and bound it with $\min(\mathcal{R})$.} until all tiles are assigned with bitrates, where $scale$ denotes the scaling step. The rationale behind is that tiles distant from the predicted viewport are of lower viewing probability and therefore can be allocated with lower bitrates for bandwidth efficiency. 

\begin{figure}[t]
    \centering
    \begin{minipage}[t]{0.46\textwidth}
    \subfigure[Agent]{
    \includegraphics[width=\textwidth]{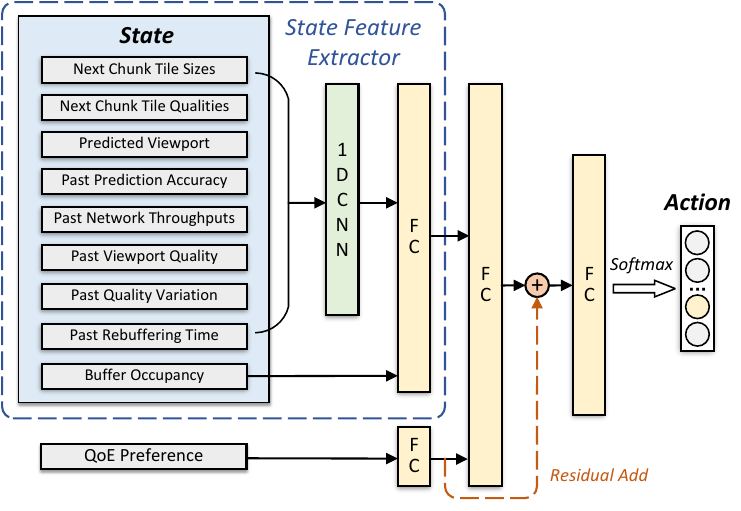}
    \label{subfig:agent_dnn}
    }
    \end{minipage}
    \begin{minipage}[t]{0.47\textwidth}
    \subfigure[QoE identifier]{
    \includegraphics[width=\textwidth]{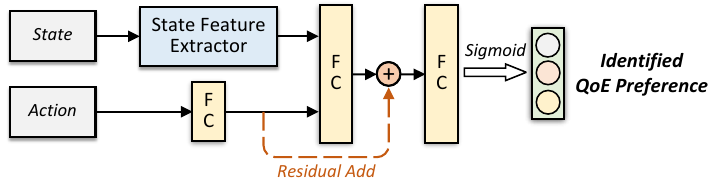}
    \label{subfig:qoe_dnn}
    }
    \end{minipage}
    % \begin{minipage}[t]{0.4\textwidth}
    % \subfigure[Discriminator]{
    % \includegraphics[width=\textwidth]{figures/disc_dnn2.pdf}
    % \label{subfig:disc_dnn}
    % }
    % \end{minipage}
    \vspace{-0.3cm}
    \caption{Neural network architectures of the agent and QoE identifier.}
    \label{fig:dnn_architecture}
    % \vspace{-0.3cm}
\end{figure}

\textbf{Network architecture.} Figure~\ref{subfig:agent_dnn} depicts the NN architecture of the agent. For vectorized information ${Z}_c, {R}_c, \vec{{v}}_c, \vec{g}_c, \vec{n}_c, \vec{q}^{\,1}_c, \vec{q}^{\,2}_c, \vec{q}^{\,3}_c$, we use 1D convolution layers to extract hidden features from each input. These features are next flattened and fed to fully-connected (FC) layers. The buffer occupancy $b_c$ and QoE preference $w$ are directly fed to FC layers for feature extraction. All hidden features are concatenated together and sequentially fed to another two FC layers to learn complex relationship between different features. Finally, a softmax layer is used to output the probability distribution of each action.

\subsection{Mutual Information-Based Reward Modeling}

The key to tackle the challenge of QoE preference diversity is to train the agent to understand the relationship between the input QoE preference and its output bitrate actions. One natural approach is to train the agent with multiple preferences with the reward defined as the QoE scores calculated with different preference weights. Nevertheless, as described in Section~\ref{subsec:motivation_abr}, this approach is prone to catastrophic forgetting and generalization issues. Moreover, even though the QoE preference is explicitly fed to the agent, without any restrictions on how such information should be utilized, the agent is free to ignore it for bitrate selection, which prevents the agent to maximize diverse QoE objectives according to the input preference.

To combat the above limitations, we augment the reward for agent training based on representation learning (RepL). RepL is the advanced technique that enables effective learning of task-specific representations that reveal meaningful patterns and alleviate the complexity of the task-solving process~\cite{bengio2013representation}. Specifically, we introduce mutual information into the reward function to facilitate the agent to learn useful representations that capture the salient attributes of users' preferences (e.g., preferences of bitrate quality and rebuffering). In information theory, the mutual information $I(X;Y)$ between variable $X$ and variable $Y$ measures the amount of information learned from $Y$ about $X$~\cite{chen2016infogan}. In our case, we expect to maximize the amount of information learned from the agent's output action $a$ about the input preference weight $w$, so that the agent can dynamically select bitrates according to the input preference. In other words, the information of $w$ should not be lost in the decision making process. Hence, the reward function to train the agent is defined as follows:
\begin{equation}
    rew_c = (1 - \alpha)QoE_c(w) + \alpha I(w;a_c,s_c)
\end{equation}
Here, $rew_c$ stands for the reward signal of downloading chunk $c$. $QoE_c(w)$ represents the QoE scores calculated with preference weight $w$ according to equation (9). $I(w;a_c,s_c)$ denotes the mutual information between the input preference $w$ and agent's output action $a_c$ selected when encountering state $s_c$. Finally, $\alpha\in[0,1]$ is the weight parameter to control the trade-off between the two components.
% where $\pi$ represent the agent policy, $\pi(a_c|s_c,w_i)$ represent the action selected by the agent under state $s_c$ and preference $w_i$, and $\alpha$ is the weight parameter to control the trade-off between the two components.

The technical challenge of implementing mutual information lies in the difficulty of exact computation~\cite{belghazi2018mutual}. Fortunately, recent advances have showcased the effective estimation of mutual information using NN models~\cite{chen2016infogan}\cite{li2017infogail}\cite{hjelm2018learning}. Therefore, we design QoE identifier, an efficient model to estimate mutual information. As demonstrated in Figure~\ref{fig:training_framework}, the QoE identifier is essentially a neural regressor that identifiers the information of preference weight $w$ from the action $a_c$ that agent takes when encountering state $s_c$. It takes a state-action pair from the agent as input and outputs the identified QoE preference, which will be used for efficient estimation of mutual information. With the QoE identifier, the reward for training the agent is modeled as:
\begin{equation}
    \setlength{\abovedisplayskip}{3pt}
    rew_c=(1-\alpha)QoE_c(w)-\alpha\log MSE(w;Q_\delta(s_c,a_c))
    \setlength{\abovedisplayskip}{3pt}
\end{equation}
where $Q_\delta$ represents the NN model of QoE identifier parameterized by $\delta$, and $MSE(w;Q_\delta(s_c,a_c))$ denotes the mean square error between the true preference weight $w$ and the one identified by QoE identifier $Q_\delta(s_c,a_c)$. 
The term $-\log MSE(w;Q_\delta(s_c,a_c))$ quantifies the amount of mutual information between the QoE preference and the selected action. The higher of its value, the greater similarity between true preference $w$ and the identified one $Q_\delta(s_c,a_c)$, thereby indicating higher mutual information. The rationale behind is that if the agent’s bitrate decisions capture the QoE preference $w$ properly, the QoE identifier should be able to extract the information of $w$ from its state $s_c$ and action $a_c$, and vice verse. Therefore, through reward optimization, our agent is capable to adaptively maximize diverse QoE objectives based on the users' preferences.

Figure~\ref{subfig:qoe_dnn} illsutrates the architecture of the QoE identifier model. We design the model with the same state feature extractor as the agent. Besides, we process the action one-hot vector with a FC layer, pass the features sequentially to another two FC layers, and use sigmoid as the output layer. To be consistent to the number of QoE preference weight parameters, the number of output neurons is set to 3. It is worth noting that the QoE identifier is designed to guide the training of agent, and therefore is used only in the training phase.
% As shown in Figure~\ref{subfig:disc_qoe_dnn}, we design the QoE identifier with the same architecture to the discriminator except that that the number of output neurons is 3 to be consistent to the number of QoE weight parameters. 

\begin{algorithm}[t]
\caption{Training procedure of RepL-based bitrate selection algorithm}
\label{algo:br_training}
\begin{algorithmic}[1] %这个1 表示每一行都显示数字
    \REQUIRE {QoE preference pool $\mathcal{W}$; weight parameter $\alpha$}.
    \ENSURE {Learned agent $\pi_\theta$.}
    \STATE Initialize the parameters of agent as $\pi_{\theta_0}$.\\
    \STATE Initialize the parameters of QoE identifier as $Q_{\delta_0}$.\\
    \FOR{$i = 0, 1, 2, \cdots$} 
        \STATE Sample a batch of QoE preferences: $w_i \sim \mathcal{W}$. \\
        \STATE Sample trajectories: $\tau_i\sim \pi_{\theta_i}(w_i)$, with the preference weight fixed during each rollout.\\
        \STATE Update $\delta_i \rightarrow \delta_{i+1}$ by descending with gradients:
        \begin{equation}
            % \Delta_{\delta_i} = \alpha\mathbb{E}_{\tau_i}[\nabla\log\rho(w; I_{\delta_{i}}(s,a))] \notag
            \Delta_{\delta_i} = \mathbb{E}_{(s, a)\sim\tau_i}[\nabla_{\delta_i} MSE(w_i; Q_{\delta_{i}}(s,a))] \notag
        \end{equation} \\
        \STATE Take a policy update step from $\theta_i$ to $\theta_{i+1}$ using the PPO update rule to optimize reward:
        \begin{equation}
            rew = (1-\alpha)QoE(w_i) - \alpha\log MSE(w_i; Q_{\delta_{i+1}}(s,a)) \notag
        \end{equation}
    \ENDFOR
\end{algorithmic}
\end{algorithm}

\subsection{Training Methodology}
\label{subsec:bitrate_selection_training_algorithm}

We introduce adversarial training to update the parameters of agent and QoE identifier. Let $\pi_\theta$ and $Q_\delta$ represent the agent parameterized by $\theta$ and QoE identifier parameterized by $\delta$, respectively. At each training step, we sample a batch of preference weights from the weight pool: $ w_i\sim\mathcal{W}$. Then we sample the state-action trajectories under the agent policy with the preference weight fixed during each rollout: $\tau_i\sim\pi_{\theta_i}(w_i)$. Next, we update the QoE identifier model in order to calculate $Q_{\delta}(s,a)$ in equation (11) properly. The goal of QoE identifier is to optimize the mutual information between the agent's state-action trajectory and the associated QoE preference to be maximum, which is equivalent to minimize the mean square error between the true QoE preference and its identified one. Hence, its parameters can be updated $\delta_i \rightarrow \delta_{i+1}$ through the following gradients:
\begin{equation}
    \setlength{\abovedisplayskip}{3pt}
    % \Delta_{\delta} = \alpha\mathbb{E}_{\tau}[\log\rho(w; I_\delta(s,a))]
    \Delta_{\delta_i} = \mathbb{E}_{(s,a)\sim\tau_i}[\nabla_{\delta_i}  MSE(w_i; Q_{\delta_i}(s,a))]
    \setlength{\belowdisplayskip}{3pt}
\end{equation} 
Afterwards, we update the agent $\theta_i \rightarrow \theta_{i+1}$ using common reinforcement learning (RL) framework to optimize the reward calculated by (11). In this paper, we choose Proximal Policy Gradient (PPO)~\cite{schulman2017proximal} as the RL framework to update the agent's parameters. The overall training procedure is summarized in Algorithm~\ref{algo:br_training}.

\textbf{Residual learning enhancement.} During the practical implementation of the proposed RepL-based bitrate selection algorithm, we find that both the agent and QoE identifier face convergence difficulties (see Section~\ref{subsec:evaluation_bitrate_selection}). Empirically, we observe that when fusing QoE preference and state features in the agent model, the dominance of state features hinders the propagation of preference features within the network. A similar phenomenon occurs with the action features in the QoE identifier model. Consequently, the agent struggles to capture QoE preference information, while the QoE identifier fails to effectively identify the preference information from the agent's actions. To address this issue, we employ residual learning~\cite{he2016deep} to facilitate the propagation of these crucial features within the networks. Specifically, as depicted in Figure~\ref{fig:dnn_architecture}, we add the preference features and action features with the outputs of the penultimate FC layer for the agent and QoE identifier, respectively.

\begin{figure*}[t]
    \centering
    \subfigure{
        \includegraphics[width=0.42\textwidth]{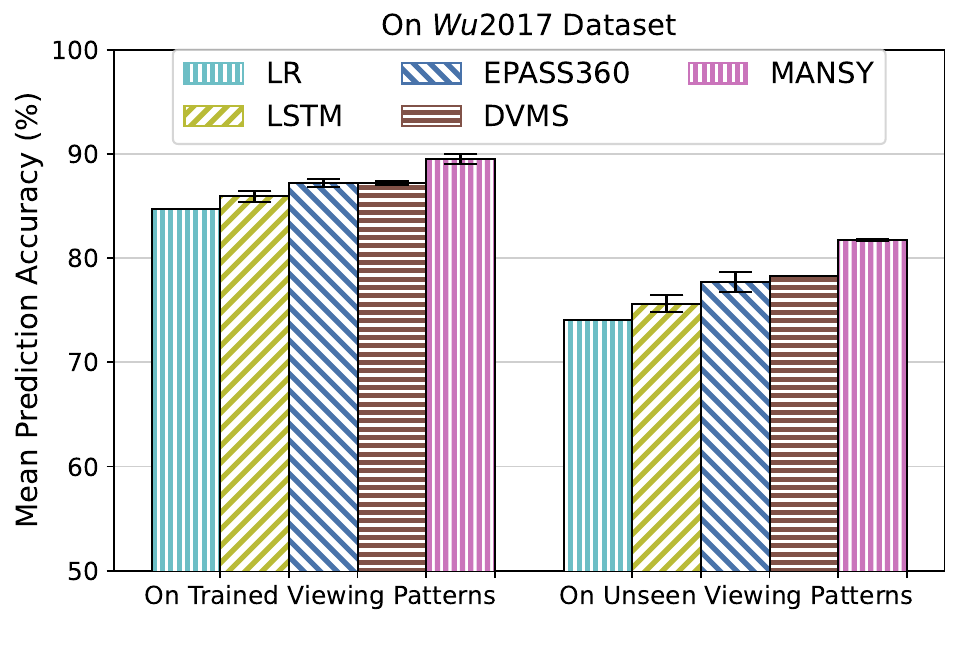}
        \label{subfig:mean_acc_cmp_seen}
        % \vspace{-0.2cm}
    }
    \hspace{0.4cm}
    \subfigure{
        \includegraphics[width=0.42\textwidth]{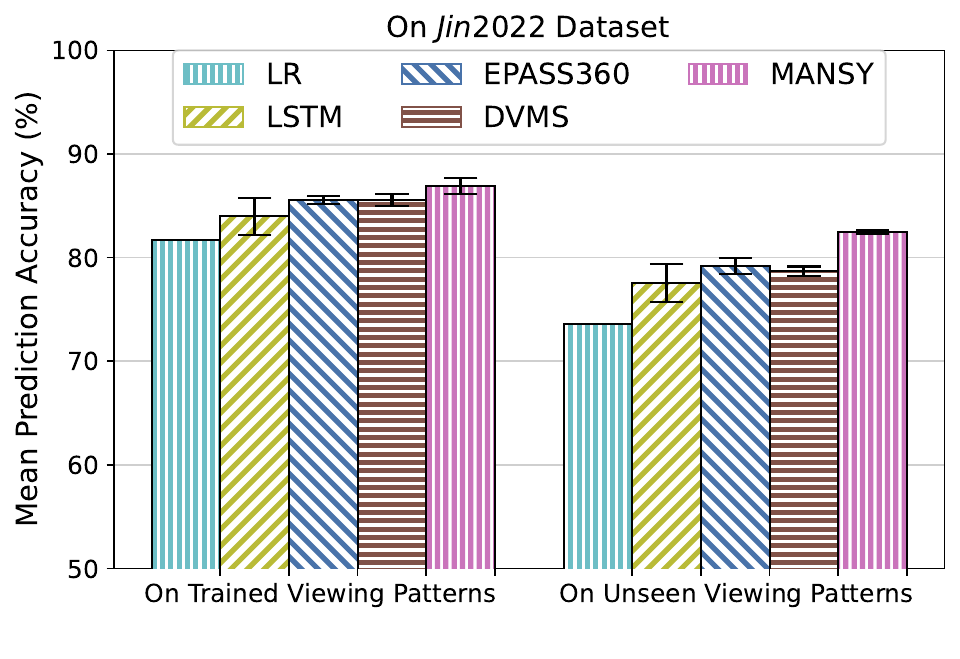}
        \label{subfig:mean_acc_cmp_unseen}
    }
    \vspace{-0.4cm}    
    \caption{\SecRevision{Mean prediction accuracy of different methods on trained and unseen viewing patterns. We show the mean and standard deviation across 3 runs with different random seeds for training and testing.}}
    \label{fig:mean_accuracy}
    % \hspace{0.1cm}
\end{figure*}

\section{Evaluation}
\label{sec:evaluation}
\subsection{Experiment Setup}

\textbf{Datasets.} We consider two large-scale immersive video datasets with viewport movement traces for evaluation, which contains numerous videos of various types watched by a large number of users:
\begin{itemize}
    \item \textit{Wu2017}~\cite{wu2017dataset}: In this dataset, 8 videos with an average length of 242 seconds are used for evaluation, with 6 for training, 1 for validation and 1 for testing. Each video in the dataset contains 48 viewport trajectories of 48 users.
    \item \textit{Jin2022}~\cite{jin2022you}: We employ 24 60-second videos watched by 60 users from this dataset. We select 18 videos for training, 3 for validation and 3 for testing.
\end{itemize}
The viewport positions of these datasets are transformed into equirectangular format according to the method in~\cite{nguyen2019saliency}. Each video is segmented into chunks of 1 second, and each chunk is further divided into 8x8 tiles. We use FFMPEG with encoder X.264 to encode the videos into $|\mathcal{R}|=5$ bitrate versions: 1Mbps (360p), 5Mbps (720p), 8Mbps(1080p), 16Mbps(2K), 35Mbps (4K). For bandwidth dataset, we consider a public dataset~\cite{hooft2016http} to simulate real-world network conditions, which contains 40 bandwidth traces with various fluctuation patterns. We select 24 traces for training, 8 for validaiton and 8 for testing.

\textbf{Baselines.} We compare our approach with other state-of-the-art methods. Specifically, for viewport prediction, we implement the following baselines for comparison:
% For viewport prediction, we compare \textit{MANSY} with the following SOTA prediction methods.
\begin{itemize}
    \item \SecRevision{\textit{LR}~\cite{xie2017360probdash} employs a simple linear regression model to predict the trends of viewport movements.}
    \item \textit{LSTM}~\cite{zhang2019drl360} trains a basic LSTM model for viewport prediction.
    \item \textit{EPASS360}~\cite{zhang2020epass360} is similar to \textit{LSTM}, except that it leverages model ensemble to explicitly set up three LSTM models and ensembles their prediction results.
    \item \textit{DVMS}~\cite{guimard2022deep} designs a Gate Recurrent Unit (GRU) model based on variational auto-encoder (VAE), which predicts five viewport trajectories to capture the variation of users' viewing patterns at the cost of multiple forward passes during inference phase.
\end{itemize}
For bitrate selection, the following baselines are implemented:
\begin{itemize}
    \item \textit{PARIMA}~\cite{chopra2021parima} employs a heuristic algorithm that allocates bitrates to tiles according to their distance to the predicted viewports based on the estimated bandwidth.
    \item \textit{RLVA}~\cite{jiang2020reinforcement} designs a DRL algorithm for bitrate selection. It trains the DRL agent to optimize only one single QoE preference.
    \item \textit{PAAS}~\cite{wu2021paas} proposes a dynamic-preference scheme to train a DRL agent to simultaneously optimize multiple QoE preferences. It defines reward as the combination of QoE scores calculated under the current preference and the ones calculated under another randomly sampled preference, so as to alleviate forgetting problem.
    \item \SecRevision{\textit{Pensieve}~\cite{mao2017neural} is a DRL-based bitrate selection algorithm for traditional non-immersive videos, which allocates the same bitrate to all tiles for each video chunk. We implement this method to verify whether the non-immersive video streaming algorithm can work effectively in the case of immersive video streaming.}%the effectiveness of non-immersive video streaming algorithms in the case of immersive video streaming.}
\end{itemize}

\textbf{Parameter settings.} 
By default, the number of input-output heads of the proposed MTIO Transformer model is set to $M=3$. Besides, we configure $d_e=512, N_{ah}=8, d_k=d_v=64, N_{block}=2$ and the learning rate as 1e-4. For bitrate selection, we empirically feed past $k=8$ sample information into the agent. The number of filters of all 1D CNN layers is 128, and the stride is 1. Their kernel sizes are set to the length of input vectors. The size of the FC layer to extract features from QoE preference is set to 128, while the sizes of the last two FC layers are set to 1280, 128, respectively. The same setting are applied on the QoE identifier model. In addition, we configure the learning rate of the agent as 5e-4, reward discount factor as 0.95, entropy coefficient as 0.02 and weight parameter $\alpha$ as 0.5. The learning rate of the QoE identifier is configured as 1e-4.

\textbf{Metrics.} We use the prediction accuracy and QoE scores as the evaluation metrics. In particular, the prediction accuracy is measured as the intersection of union (IoU) between the predicted viewport and ground-truth viewport~\cite{rondon2021track}. \Revision{To be more specific, the prediction accuracy is calculated by:}
\begin{equation}
    \setlength{\abovedisplayskip}{3pt}
    % \Delta_{\delta} = \alpha\mathbb{E}_{\tau}[\log\rho(w; I_\delta(s,a))]
    Accuracy = \frac{FoV(x^p, y^p)\cap FoV(x^g, y^g)}{FoV(x^p, y^p)\cup FoV(x^g, y^g)} \notag
    \setlength{\belowdisplayskip}{3pt}
\end{equation} 
\Revision{where $(x^p, y^p)$ and $(x^g, y^g)$ denote the positions of the predicted and ground-truth viewport centers, respectively; $FoV(x,y)$ denotes the field of view (FoV), i.e., viewport area, centered at position $(x,y)$. According to the specifications of existing popular headsets~\cite{fov}, we configure the size of FoV to be 16\% of the total video area.}

\Revision{\textbf{Hardware settings.} We conduct all experiments on a desktop computer equipped with an Intel(R) Core(TM) i7-12700 CPU and NVIDIA 3090 GPU. Note that the computing resources to run our experiments are highly redundant.}

\subsection{Viewport Prediction}
\label{subsec:evaluate_viewport_prediction}
% In this part, we evaluate the performance of the proposed MTIO Transformer of \textit{MANSY}, on both trained and unseen viewing patterns. Note that we apply the method in~\cite{lu2022personalized} to categorize users into eight groups of diverse viewing patterns, with six groups for training and the rest regarded as unseen viewing patterns. 
In this part, we first evaluate the performance of MTIO-Transformer viewport prediction model of \texttt{MANSY}. By default, we use the viewports in the last second to predict the future viewports in the next second, i.e., $h=H=1\text{s}$. 
% Additionally, we leverage the method in~\cite{lu2022personalized} to categorize users into eight groups of diverse viewing patterns based on their viewport distribution similarities.
Additionally, we leverage the method in~\cite{lu2022personalized} to categorize users into seven groups of diverse viewing patterns based on their viewing preferences (e.g., preferring to watch dynamic objects or focus on static objects).
We select five groups for training and the rest for evaluating generalization performance. 
% We select five groups for training and consider the rest two groups as users with unseen viewing patterns.
When evaluating each method on the trained/unseen viewing patterns, we report the their performance on the trained/unseen groups on the testing videos.

\subsubsection{Comparison With Baselines}

\begin{figure*}[t]
    \centering
    \includegraphics[width=0.42\textwidth]{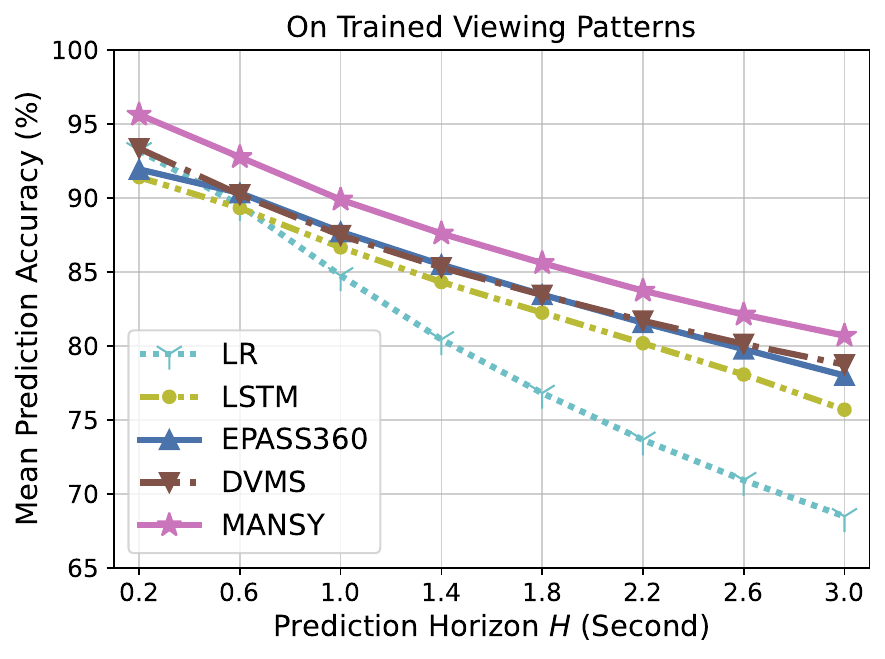}
    \hspace{0.6cm}
    \includegraphics[width=0.42\textwidth]{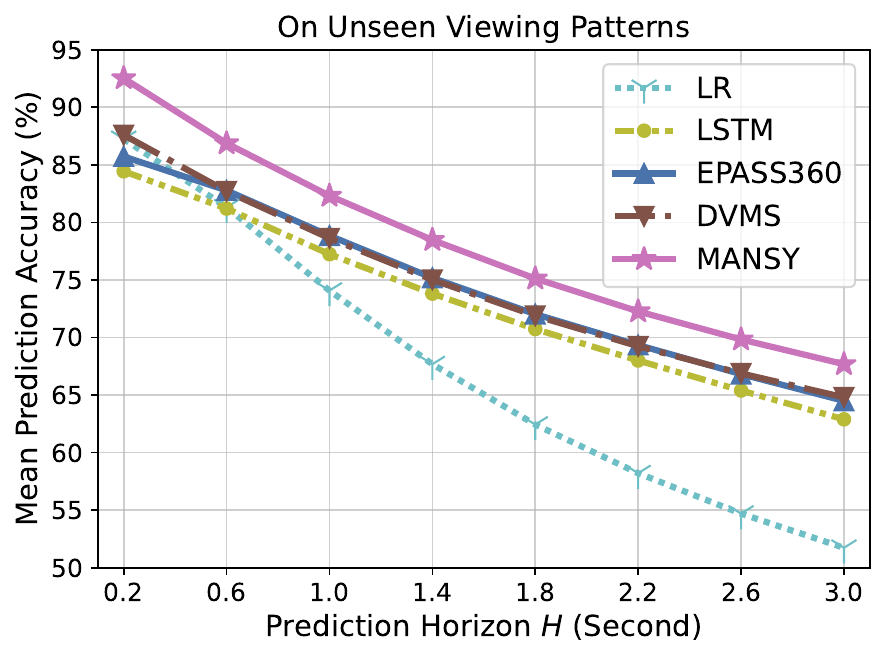}
    \vspace{-0.2cm}
    \caption{\SecRevision{Mean prediction accuracy of different methods with different prediction horizon $H$ on \textit{Wu2017} dataset.}}
    \label{fig:accuracy_vs_time}
\end{figure*}

\begin{figure}
    \centering
    \includegraphics[width=0.42\textwidth]{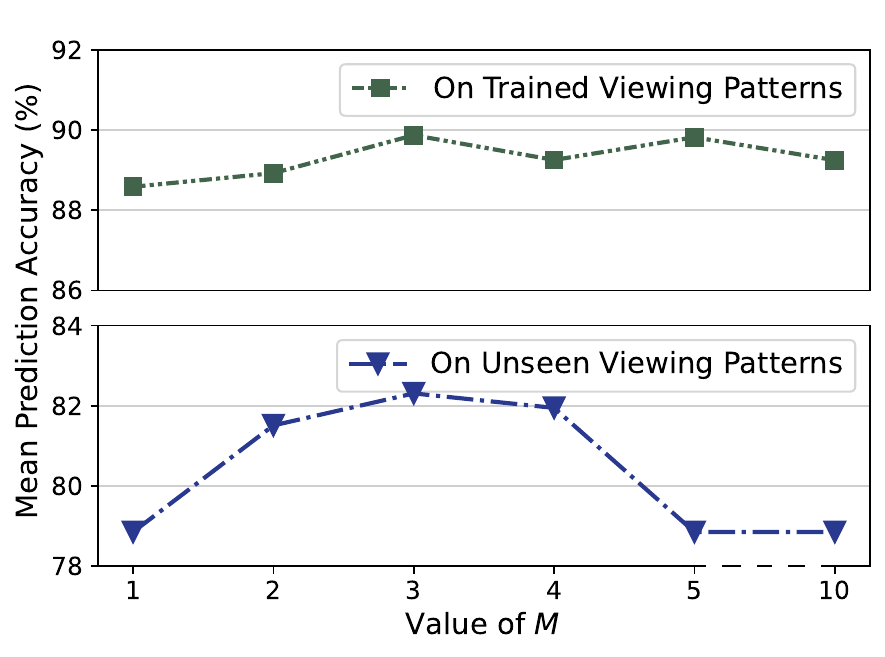}
    \vspace{-0.4cm}
    \caption{\Revision{Effects of the number of input-output heads $M$.}}
    \label{fig:effects_of_m}
\end{figure}

Figure~\ref{fig:mean_accuracy} compares the mean viewport prediction accuracy of different methods. As shown in Figure~\ref{fig:mean_accuracy}, \texttt{MANSY} outperforms other baselines on both trained and unseen viewing patterns. \Revision{On \textit{Wu2017} (\textit{Jin2022}) dataset, it effectively improves the absolute mean accuracy by 2.3\%--4.8\% and 3.4\%--7.7\% (1.3\%--5.2\% and 3.8\%--8.8\%) on trained and unseen viewing patterns, respectively.} Notably, \texttt{MANSY} shows more significant improvement on the unseen viewing patterns, which highlights its superior generalization performance. The superiority of \texttt{MANSY} can be attributed to two key aspects. 
\begin{itemize}
    \item First, \texttt{MANSY} employs the MTIO architecture to efficiently reduce the prediction bias and thus achieves better generalization performance over diverse viewing patterns. In contrast, \textit{LSTM} neglects the prediction bias towards the training data and thus experiences more drastic accuracy loss when testing on unseen viewing patterns.
    \item  Second, \texttt{MANSY} designs the prediction model based on Transformer with the attention mechanism to effectively learn long-term dependencies and predict the trends of viewport movement more accurately. By comparison, both \textit{EPASS360} and \textit{DVMS} adopt conventional LSTM or GRU models for viewport prediction, which limit their ability to capture the viewport moving patterns, thus resulting in poorer performance than \texttt{MANSY}.
\end{itemize}

We further compare the accuracy of different methods with different prediction horizon $H$ on \textit{Wu2017} dataset.
\Revision{As depicted in Figure~\ref{fig:accuracy_vs_time}, \texttt{MANSY} consistently achieves the highest prediction accuracy, with the improvements of 2.4\%--17.8\% and 4.2\%--30.8\% on trained and unseen viewing patterns, respectively.} This indicates the stronger ability of \texttt{MANSY} in both short-term and long-term prediction. The results on \textit{Jin2022} dataset are similar and thus are omitted here for brevity. 

\begin{table}[t]
    \centering
    % \begin{minipage}{0.33\textwidth}
        % \begin{flushright}
    % \flushright
    \caption{\Revision{Comparison of computation overhead of different architectures. ``$\uparrow$ x\%'' means the increase compared to standard architecture with one input-output head ($M=1$).}}
    \label{tab:complexity_comparison}
    \vspace{-0.2cm}
    \begin{tabular}{m{2.7cm}<{\centering}  m{2.6cm}<{\centering}  m{2.3cm}<{\centering} }
        \toprule
        \textbf{Architecture} & \textbf{Memory Consumption (MB)} & \textbf{Inference \;\;\;\;\;Time (ms)} \\
        % \cline{2-3}
        % & \textbf{Parameters} & \textbf{FLOPs}\\
        \midrule
        Standard ($M=1$) & 28.12 & 35.63  \\
        % MTIO (M=2M=2) & 0.28\% & 0.24\%  \\
        % \midrule
        MTIO ($M=3$) & 28.13 ($\uparrow$ 0.04\%) & 35.66 ($\uparrow$ 0.78\%)  \\
        % \midrule
        % MTIO (M=4M=4) & 0.83\% & 0.72\%  \\
        % \midrule
        MTIO ($M=5$) & 28.15 ($\uparrow$ 0.11\%) & 35.85 ($\uparrow$ 1.43\%)  \\
        % \midrule
        MTIO ($M=10$) & 28.19 ($\uparrow$ 0.25\%) & 36.35 ($\uparrow$ 1.56\%)  \\
        % \midrule
        VAE~\cite{guimard2022deep} & 32.12 ($\uparrow$ 10.67\%) & 153.54 ($\uparrow$ 430.94\%)  \\
        % \midrule
        Explicit Ensemble~\cite{zhang2020epass360} & 84.35 ($\uparrow$ 200.00\%) & 106.89 ($\uparrow$ 200.00\%)  \\
        \bottomrule
    \end{tabular}
    % \end{minipage}
\end{table}

\begin{figure*}[t]
    % \centering
    % \begin{minipage}[t]{0.9\textwidth}
    \centering
    \subfigure{
        \includegraphics[width=0.42\textwidth]{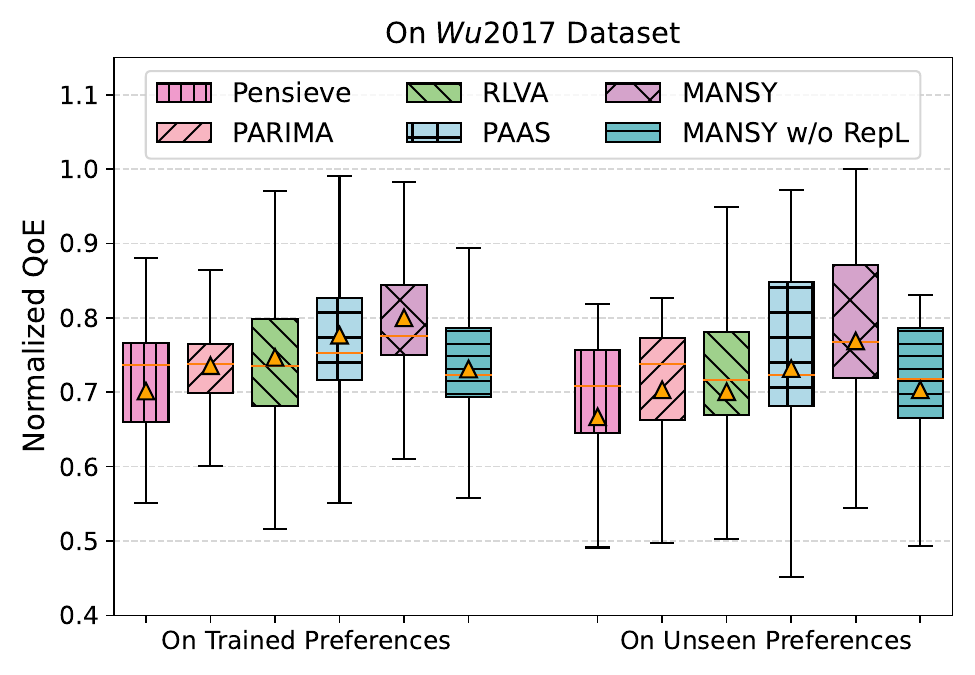}
        % \vspace{-0.2cm}
    }
    \hspace{0.4cm}
    \subfigure{
        \includegraphics[width=0.42\textwidth]{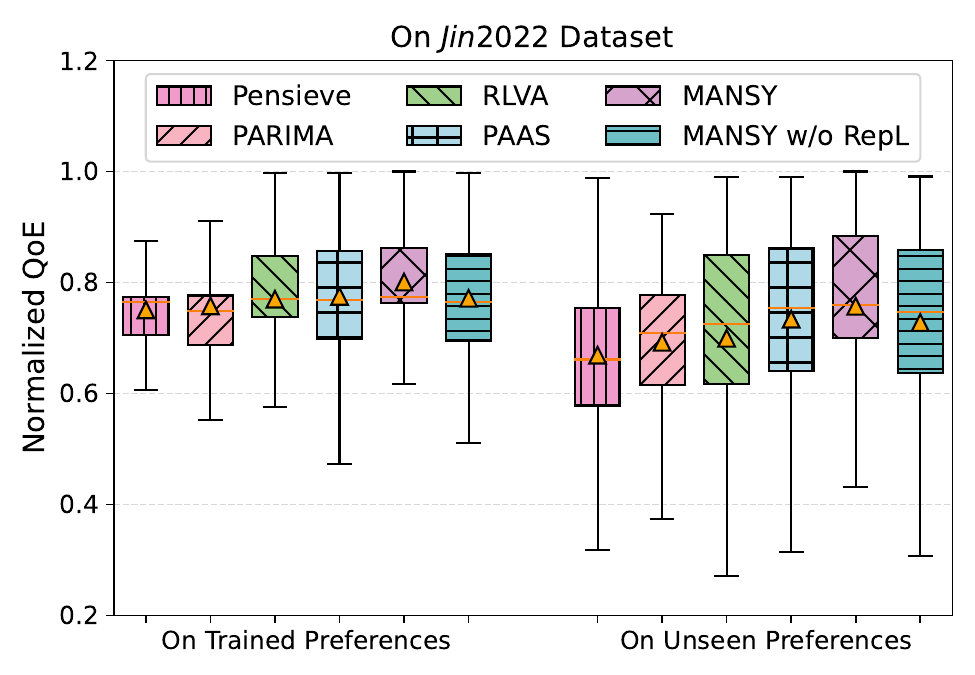}
    }
    \vspace{-0.2cm}    
    \caption{\SecRevision{Normalized QoE of different methods on trained and unseen QoE preferences. The shape of box shows the QoE distribution and the triangle in each box denotes the mean QoE.}}
    \label{fig:box_qoe}
    % \end{minipage}
\end{figure*}

\subsubsection{Effectiveness of MTIO Architecture}

Next, we evaluate the effectiveness of the proposed MTIO architecture. We report the mean accuracy of \texttt{MANSY} with different number of input-output heads $M$ in Figure~\ref{fig:effects_of_m}. Note that $M=1$ is equivalent to the standard architecture with single input-output head. As shown, increasing $M$ will improve the predictive performance especially on unseen viewing patterns, which confirms the effectiveness of MTIO architecture to reduce prediction bias and improve generalization. Besides, we also observe that such performance gain will gradually diminish when $M$ is sufficiently large (e.g., $M > 3$). 
% One possible reason is that a large $M$ will quickly reach the neural network capacity, and thus the implicitly trained sub-models may share high similarities. Consequently, their bias may accumulate, hurting the benefits of ensemble.
This phenomenon could be attributed to the fact that our MTIO architecture utilizes a single neural network to implicitly train multiple sub-models, and a large $M$ will quickly reach the network capacity. Consequently, the trained sub-models may share high similarities, while the success of ensemble learning relies on the diversity of sub-models~\cite{havasi2021training}. As a result, their prediction bias may accumulate, hurting the benefits of ensemble. This suggests that in practice, it is unnecessary to set $M$ too large for the MTIO architecture.

As a supplement, we also measure the computation overhead of MTIO architecture \Revision{in terms of the increase of the memory consumption and inference time\footnote{\Revision{Time measurement is performed on an Intel(R) Core(TM) i7-12700 CPU, and is limited to \textit{use only 1 CPU core} to simulate resource-constrained scenarios.}} compared to the standard architecture (i.e., $M=1$).} As shown in Table~\ref{tab:complexity_comparison}, the overhead introduced by MTIO is negligible: even when $M=10$, \Revision{it only increases 0.25\% of memory consumption and 1.56\% of inference time.} By comparison, when adopting VAE or explicit ensemble, as in~\cite{guimard2022deep}\cite{zhang2020epass360}, \Revision{the overhead drastically increase to 10.67\%/430.94\% or 200\%/200\% on memory consumption and inference time, respectively.}

\SecRevision{We also measure the overhead of the \textit{LR} method and results show that it consumes 2.66 MB memory and takes 11.77 ms per inference. Considering that \textit{LR} is a non-DNN-based method, it naturally causes less computation overhead than DNN-based ones, including our MTIO Transformer. Despite this, our MTIO Transformer achieves significantly higher prediction accuracy than \textit{LR}, especially for long-term prediction, as illustrated in Figure~\ref{fig:accuracy_vs_time}. Moreover, modern commercial VR headsets are now equipped with sufficient computing resources to support the exeuction of small-size DNNs. For example, Meta Quest Pro~\cite{metaquestpro} is powered by the Qualcomm Snapdragon XR2@1.8GHz, which features 8 CPU cores and 12GB memory along with an Adreno 650 GPU for accelerating computation. Hence, deploying our model on commercial headsets is feasible and the performance-overhead trade-off of our model is worthwhile.}

\subsection{Bitrate Selection}
\label{subsec:evaluation_bitrate_selection}

In this part, we evaluate the performance of \texttt{MANSY} in bitrate selection. 
Following previous work~\cite{wu2021paas}, we construct 8 QoE weights with diverse preferences\footnote{The full list of the QoE preferences is omitted here for brevity and can be found in our codes.} on different QoE metrics (e.g., high-bitrate-first and low-rebuffering-first), with 4 used for training and the rest for generalization evaluation. In the following, we report the performance of different methods on the trained/unseen QoE preferences on the testing videos and users. 
\Revision{Besides, since \textit{RLVA} follows the single-preference optimization scheme, we follow the idea in~\cite{zhang2019drl360}\cite{lu2022personalized}\cite{wang2017user} to train personalized \textit{RLVA} models for each QoE preference separately. 
When it comes to an unseen preference, we use the \textit{RLVA} model trained with the QoE preference of the highest cosine similarity for testing.} \SecRevision{We adopt the same strategy for \textit{Pensieve}, the single-preference algorithm originally designed for non-immersive video streaming.}
% When testing on unseen QoE requirements, we use the model trained with QoE requirement that is the most similar to the specific unseen QoE requirement. The similarity between two QoE requirements is measured as the cosine similarity between their QoE weights.

\begin{figure*}[t]
    \centering
    \subfigure[On trained preference $(\frac{7}{9}, \frac{1}{9}, \frac{1}{9})$]{
        \includegraphics[width=0.42\textwidth]{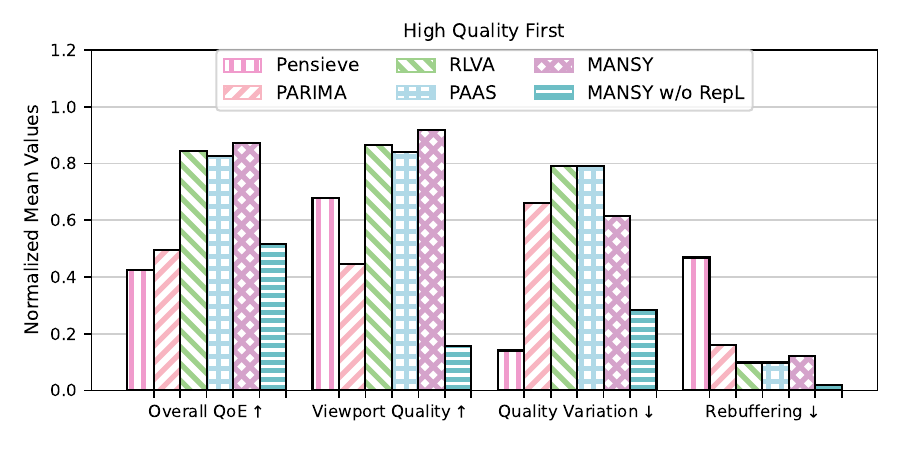}
        \label{subfig:upper_left}
    }
    \hspace{0.3cm}
    \subfigure[On trained preference $(\frac{1}{9}, \frac{1}{9}, \frac{7}{9})$]{
        \includegraphics[width=0.42\textwidth]{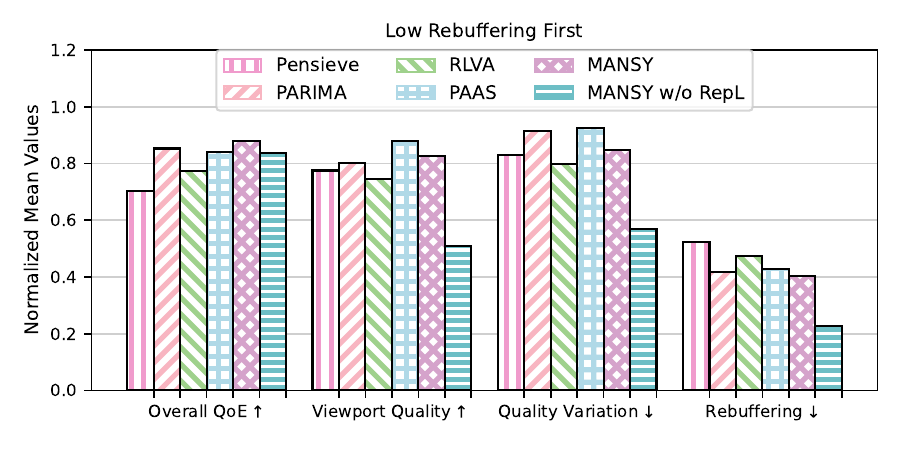}
        \label{subfig:upper_right}
    }
    \subfigure[On unseen preference $(\frac{5}{9}, \frac{1}{3}, \frac{1}{9})$]{
        \includegraphics[width=0.42\textwidth]{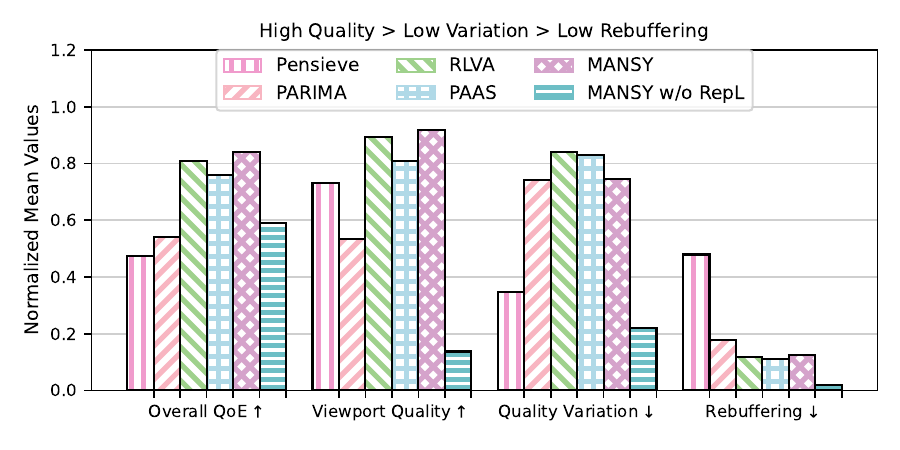}
        \label{subfig:lower_left}
    }
    \hspace{0.3cm}
    \subfigure[On unseen preference $(\frac{1}{9}, \frac{5}{9}, \frac{1}{3})$]{
        \includegraphics[width=0.42\textwidth]{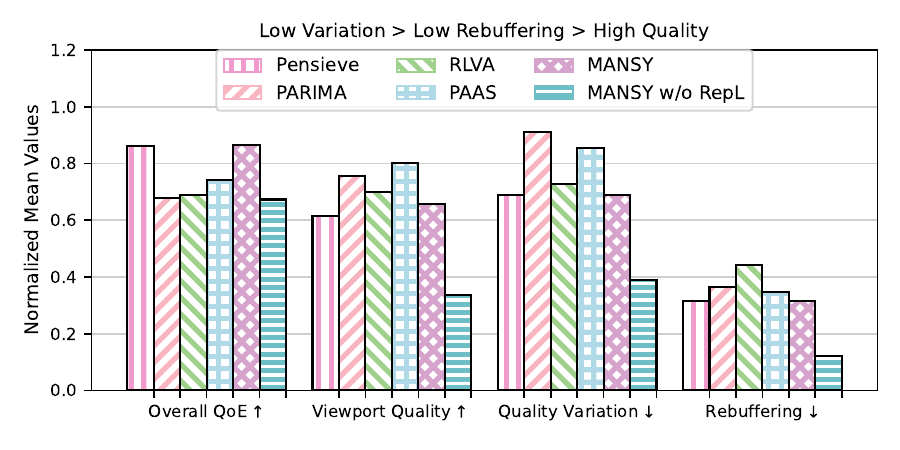}
        \label{subfig:lower_right}
    }
    \vspace{-0.1cm}
    \caption{\SecRevision{Normalized mean values of different QoE metrics of different methods on various preferences on \textit{Wu2017} dataset. The arrow $\uparrow$ / $\downarrow$ means higher/lower is better.}}
    \label{fig:individual_qoe}
\end{figure*}

\subsubsection{Comparison With Baselines}

Figure~\ref{fig:box_qoe} compares the performance of different methods on different sets of preferences and datasets in terms of mean QoE and QoE distribution. As shown, \texttt{MANSY} outperforms other methods in different cases. Specifically, on \textit{Wu2017} (\textit{Jin2022}) dataset, compared to \SecRevision{\textit{Pensieve}, \textit{PARIMA}, \textit{RLVA} and \textit{PAAS}, \texttt{MANSY} improves the average QoE by 14.1\%, 9.2\%, 7.4\%, 3.1\% and 15.3\%, 9.3\%, 9.7\%, 5.1\% (6.7\%, 6.0\%, 4.3\%, 3.5\% and 13.2\%, 10.2\%, 8.9\%, 3.4\%) on trained and unseen QoE preferences, respectively.} In particular, \texttt{MANSY} generally demonstrates more significant improvement on unseen preferences, thus achieving stronger generalization ability. Besides, a large proportion of QoE values of \texttt{MANSY} is concentrated in the larger range, which further demonstrates the superiority of \texttt{MANSY}. 
% The performance gain of \texttt{MANSY} stems from the usage of the advanced RepL technique to train the DRL agent. It designs a QoE identifier to encourage the agent to maximize the mutual information between QoE preferences and bitrate decisions, enabling the agent to generalize across diverse QoE preferences. 
The performance gain of \texttt{MANSY} stems from the design of QoE identifier based on RepL to train the agent to maximize the mutual information between QoE preferences and bitrate decisions, enabling the agent to generalize across diverse QoE preferences. 

To gain a comprehensive understanding of the performance of each method, Figure~\ref{fig:individual_qoe} presents the mean values of different QoE metrics of different methods across various preferences.
\SecRevision{From Figure~\ref{fig:individual_qoe}, we can see that \textit{Pensieve} consistently yields low quality variation across all cases and matches \texttt{MANSY}'s performance for preferences that emphasize low quality variation (e.g., Figure ~\ref{subfig:lower_right}). This is because \textit{Pensieve} allocates the same bitrates for all tiles of the video chunks and thus removes intra-variation of viewport quality. However, such strategy will allocate unnecessarily high bitrates to tiles outside viewports. Therefore, \textit{Pensieve} consistently yields low viewport quality or high rebuffering time (e.g., Figure~\ref{subfig:upper_right} and Figure~\ref{subfig:lower_left}), as tiles outside viewports could be allocated with lower bitrates to reserve bandwidth for improving the tile quality inside viewports or reducing buffering time. In consequence, \textit{Pensieve} generally achieves lower QoE than other immersive video streaming algorithms in most cases, as depicted in both Figure~\ref{fig:box_qoe} and Figure~\ref{fig:individual_qoe}. This indicates that the non-immersive video streaming algorithm is not well suited for immersive videos}

As a heuristic algorithm, \textit{PARIMA} allocates bitrates to tiles according to the estimated bandwidth. However, due to the underestimation of bandwidth, \textit{PARIMA} tends to conservatively select low bitrates to prevent rebuffering. %, which leads to underutilization  of the available bandwidth resources. 
Consequently, it achieves comparable performance to \texttt{MANSY} only on preferences that emphasize low rebuffering (e.g., Figure~\ref{subfig:upper_right}), while significantly underperforming in other cases. 
% For \textit{RLVA}, we observe that when the unseen preference is similar to the one it optimizes during training, it may achieve satisfactory performance (e.g., Figure~????????????\ref{subfig:upper_left} and Figure~????????????\ref{subfig:lower_left}).
For \textit{RLVA}, although it may achieve satisfactory performance on the unseen preference similar to the one optimized during training (e.g., Figure~\ref{subfig:upper_left} and Figure~\ref{subfig:lower_left}), it fails to accurately capture user's QoE preference for bitrate selection. For instance, as shown in Figure~\ref{subfig:lower_left}, \textit{RLVA} trained to aggressively optimize viewport quality adopts a similar bitrate selection strategy for users with requirement on maintaining low variation. In consequence, it exhibits the largest quality variation, leading to inferior performance compared to \texttt{MANSY}.

Despite incorporating a dynamic-preference scheme to train the agent for optimizing different QoE preferences, we observe that \textit{PAAS} still suffers from the forgetting problem. For example, while \textit{PAAS} outperforms \textit{RLVA} on the trained low-rebuffering-first preference (Figure~\ref{subfig:upper_right}), it demonstrates inferior performance compared to \textit{RLVA} on another trained high-quality-first preference (Figure~\ref{subfig:upper_left}). Moreover, \textit{PAAS} also exhibits limited generalization ability, as it may perform worse than \textit{RLVA} in some cases even on the unseen preferences (e.g., Figure~\ref{subfig:lower_left}). 
In contrast, thanks to the proposed RepL-based training scheme, \texttt{MANSY} successfully learns useful representations that capture the essential characteristics of users' QoE preferences, enabling dynamic bitrate selection based on input preferences. Notably, the learned representations are generalizable, empowering the agent to optimize unseen preferences (e.g., Figure~\ref{subfig:lower_right}). As a result, \texttt{MANSY} demonstrates  the most promising performance across all cases.

\subsubsection{Ablation Study}
In this part, we set up several experiments to provide a thorough understanding of the bitrate selection model of \texttt{MANSY}, including effectiveness of RepL-based training and residual learning.

\textbf{Effectiveness of RepL.} We remove the proposed RepL training scheme from \texttt{MANSY} to explore its contributions to the bitrate selection performance of \texttt{MANSY}. \Revision{Specifically, we simply train the agent with multiple QoE preferences simultaneously but without the guidance of the QoE identifier, which forms the method of \textit{MANSY w/o RepL}.}

\Revision{Figure~\ref{fig:box_qoe} illustrates the performance of \textit{MANSY w/o RepL} on both trained and unseen viewing patterns on each dataset. It can be seen that without RepL, the performance of \textit{MANSY w/o RepL} significantly decreases and even becomes worse than baselines. To gain  further insights, we analyze its performance on individual QoE preferences in Figure~\ref{fig:individual_qoe}. As depicted in Figure~\ref{fig:individual_qoe}, \textit{MANSY w/o RepL} consistently maintains the lowest rebuffering time on all cases. This suggests that without the guidance of QoE identifier to learn informative representations of users’ QoE preferences, \textit{MANSY w/o RepL} heavily suffers from the \textit{catastrophic forgetting} problem. The dominance of knowledge learned to optimize low rebuffering forces \textit{MANSY w/o RepL} to aggressively select low bitrates regardless of users’ actual QoE preferences. As a result, it only achieves better or competitive performance than baselines on the low-rebuffering preference (e.g., Figure~\ref{fig:individual_qoe} (b)), but achieves inferior performance on other ones (e.g., Figure~\ref{fig:individual_qoe} (c)). This finding further supports our previous observations\footnote{\Revision{Recall that in Section~\ref{subsec:motivation_abr}, we also train \textit{RLVA} with mulitple QoE preferences and it suffers from the similar catastrophic forgetting and generalization problems.}} in Section~\ref{subsec:motivation_abr} that simply training with multiple QoE preferences will suffer from severe catastrophic forgetting issues and lead to poor generalization performance.}  

\Revision{In contrast, \texttt{MANSY} significantly outperforms \textit{MANSY w/o RepL} and other baselines across all cases. This outcome demonstrates that \texttt{MANSY} efficiently tackles the challenge of catastrophic forgetting and achieves strong generalization performance, which confirms the effectiveness of the proposed RepL-based training scheme.}

\begin{figure}[t]
    \centering
    \begin{minipage}[t]{0.45\textwidth}
    \centering
    \includegraphics[width=0.94\textwidth]{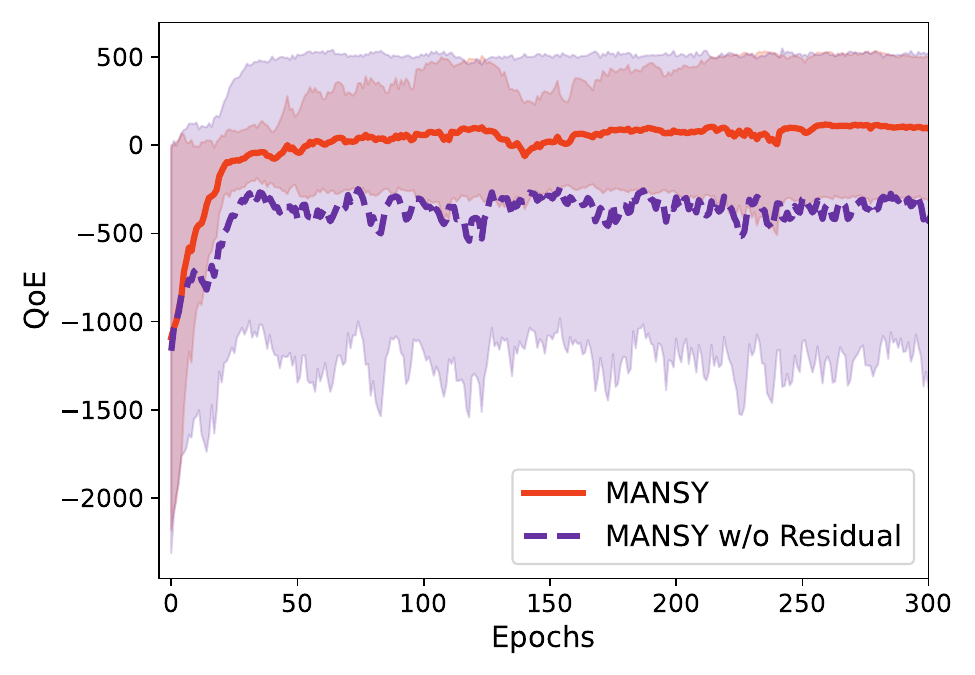}
    \vspace{-0.2cm}
    \caption{Convergence comparison of \texttt{MANSY} with/without residual learning.}
    \label{fig:convergence_cmp}
    \end{minipage}
\end{figure}

\textbf{Effectiveness of residual learning.} We next compare the convergence performance of \texttt{MANSY} and \textit{MANSY w/o Residual} to validate the effectiveness of residual learning. To be more specific, the NN architectures of the agent and QoE identifier of \textit{MANSY w/o Residual} are designed without the residual layers depicted in Figure~\ref{fig:dnn_architecture}. As illustrated in Figure~\ref{fig:convergence_cmp}, \textit{MANSY w/o Residual} faces issues related to training instability and poor asymptotic performance (i.e., final performance after convergence~\cite{xia2022genet}). This is attributed to the fact that the important features of QoE preference are lost during the fusion with state features, which make \textit{MANSY w/o Residual} difficult to capture the QoE preference information. In contrast, \texttt{MANSY} efficiently addresses this issue by adding residual layers into the NN models to facilitate the propagation of these critical features within the models. Hence, \texttt{MANSY} achieves better training stability and asymptotic performance.

\section{Related Work}
\label{sec:related_work}

\textbf{Viewport prediction.} 
% The diversity of users' viewing patterns imposes great challenges to learn-based viewport prediction methods.
As one of the main building block of tile-based immersive video streaming, the design of viewport prediction models has been extensively studied~\cite{jin2023ebublio}\cite{li2023spherical}.
% Viewport prediction~\cite{shi2022sophon}\cite{lu2022personalized}\cite{jin2023ebublio} serves as a key enabling building block of tile-based 360$^\circ$ video streaming. 
Considering the diversity of users' viewing patterns, recent studies have explored several approaches to enhance the viewport prediction models to resolve the diversity challenge. For instance, the works in ~\cite{lu2022personalized}\cite{wang2017user} group users with similar viewing patterns and train separate models for each group. The major limitation of this approach is that each time a new user group emerges, a model retraining process is required, resulting in prohibitive training cost. Alternatively, some researchers attempt to improve the generalization of viewport prediction models to serve a broad range of users. For instance, Guimard et al.~\cite{guimard2022deep} design a variational auto-encoder model that predicts multiple viewports to capture the variation of user's viewing patterns. Zhang et al.~\cite{zhang2020epass360} explicitly train several LSTM models and ensemble their prediction results to yield calibrated predictions. These methods, however, require model duplication or multiple forward passes during inference phase, leading to significant computation cost. By comparison, \texttt{MANSY} designs an efficient MTIO-Transformer model based on implicit ensemble learning. It can obtain well-calibrated predicted viewports with a single model and a single forward pass, thus improving generalization with negligible overhead.  %Note that our work differs from~\cite{zhang2020epass360} in that the proposed MTIO-Transformer implicitly trains multiple models without model duplication, thus both memory and computation overhead are greatly reduced.

\textbf{Bitrate selection.}
Recent years have witnessed the successful applications of deep reinforcement learning (DRL)
in bitrate selection of tile-based streaming~\cite{jiang2020reinforcement}\cite{kan2022rart360}\cite{li2023achieving}. However, the performance of DRL-based methods is restricted in real-world conditions where users exhibit high diversity on QoE preferences. To tackle this challenge, previous works~\cite{zhang2019drl360}\cite{kan2022rart360} propose to train different bitrate selection agents with different QoE preferences, which, however, results in prohibitive training cost and lacks scalability to adapt to new preferences. Li et al.~\cite{li2023achieving} designs a muli-agent DRL solution to allocate bitrates to users that watch the same video and share the same bottleneck link while considering their QoE preferences. Yet, their solution is built upon the strict assumption of fixed bandwidth of the bottleneck link, which often does not hold true in practice and thus limits its performance in real-world scenarios. On the other hand, Wu et al.~\cite{wu2021paas} design a dynamic-preference scheme to train the DRL agent with multiple QoE preferences simultaneously. Nevertheless, this approach suffers from the catastrophic forgetting problem, failing to generalize across diverse QoE preferences. In contrast, \texttt{MANSY} leverages the advanced RepL technique to train the agent without any specific presumptions. It designs an efficient QoE identifier to encourage the agent to automatically extract useful knowledge from users’ QoE preferences, thus generalizing across users with diverse preferences.

\section{\Revision{Discussion}}
\label{sec:discussion}
\Revision{In this section, we discuss the potential methodologies to further enhance the performance of \texttt{MANSY}. Regarding viewport prediction, the current implementation of the MTIO-Transformer in \texttt{MANSY} employs a simple averaging strategy to combine the prediction results from different input-output heads. In other words, each head is treated equally and assigned with the same weight for combining prediction. To further improve the prediction accuracy of the model, one potential approach is to introduce an adaptive weighting scheme. To be specific, each head is initially assigned with the same weight. As  the model makes prediction over time, the assigned weights can be dynamically adjusted based on the historical accuracy of each head. The underlying rationale is that a head with higher accuracy indicates that it captures user's viewing pattern more accurately, and thus should be allocated with a higher weight.}

\Revision{As for bitrate selection, \texttt{MANSY} currently focuses on optimizing the streaming services when users' QoE preferences are known in advance. To unleash its full potential, \texttt{MANSY} can be complemented with the existing active research on inferring users' QoE preferences in practice~\cite{li2022apprenticeship}\cite{zhang2021sensei}. 
One viable approach is crowdsourcing~\cite{zhang2023vidplat}, where users are invited to participate in questionnaires to gather information about their specific video watching preferences (e.g., priority of video quality).
Alternatively, users' QoE preferences can also be inferred from their historical video watching behaviors (e.g., manual bitrate switching frequency)~\cite{li2022apprenticeship}\cite{gao2018optimizing}. For example, Li et al.~\cite{li2022apprenticeship} construct a behavior dataset by collecting users' behaviors of video watching. During the online service phase, users' behaviors will be collected as they watch the videos, and a Bayes' theorem based method is utilized to calculate their QoE preferences based on the collected behaviors and constructed dataset. The above approaches can be integrated into \texttt{MANSY} to efficiently infer users' preferences to optimize bitrate selection.}

\section{Conclusion}
\label{sec:conclusion}

In this paper, we propose \texttt{MANSY}, a novel tile-based immersive video streaming system that fully captures user diversity to improve generalization. \texttt{MANSY} incorporates a Transformer-based viewport prediction model with an efficient Multi-viewport Trajectory Input Output (MTIO) architecture to reduce the prediction bias, so that it can generalize across users with diverse viewing patterns. 
% For bitrate selection, \textit{MANSY} combines the advanced Representation Learning (RepL) and Imitation Learning (IL) framework to train the agent to maxmize users' QoE with diverse requirements. 
For bitrate selection, to accommodate users' diverse QoE preferences, \texttt{MANSY} leverages representation learning (RepL) to encourage the DRL agent to learn useful representations of users' preferences by augmenting the reward function with mutual information. Considering the difficulty of exact computation of mutual information, it designs an efficient NN model called QoE identifier to estimate mutual information for reward calculation. Extensive experiments with real-world datasets confirm the superiority of \texttt{MANSY} in viewport prediction and bitrate selection on both trained and unseen viewing patterns and QoE preferences.

% % use section* for acknowledgment
% \ifCLASSOPTIONcompsoc
%   % The Computer Society usually uses the plural form
%   \section*{Acknowledgments}
% \else
%   % regular IEEE prefers the singular form
% \fi


\ifCLASSOPTIONcaptionsoff
  \newpage
\fi


\bibliographystyle{IEEEtran}
\bibliography{IEEEabrv,references}
\begin{IEEEbiography}[{\includegraphics[width=1in,height=1.25in,clip,keepaspectratio]{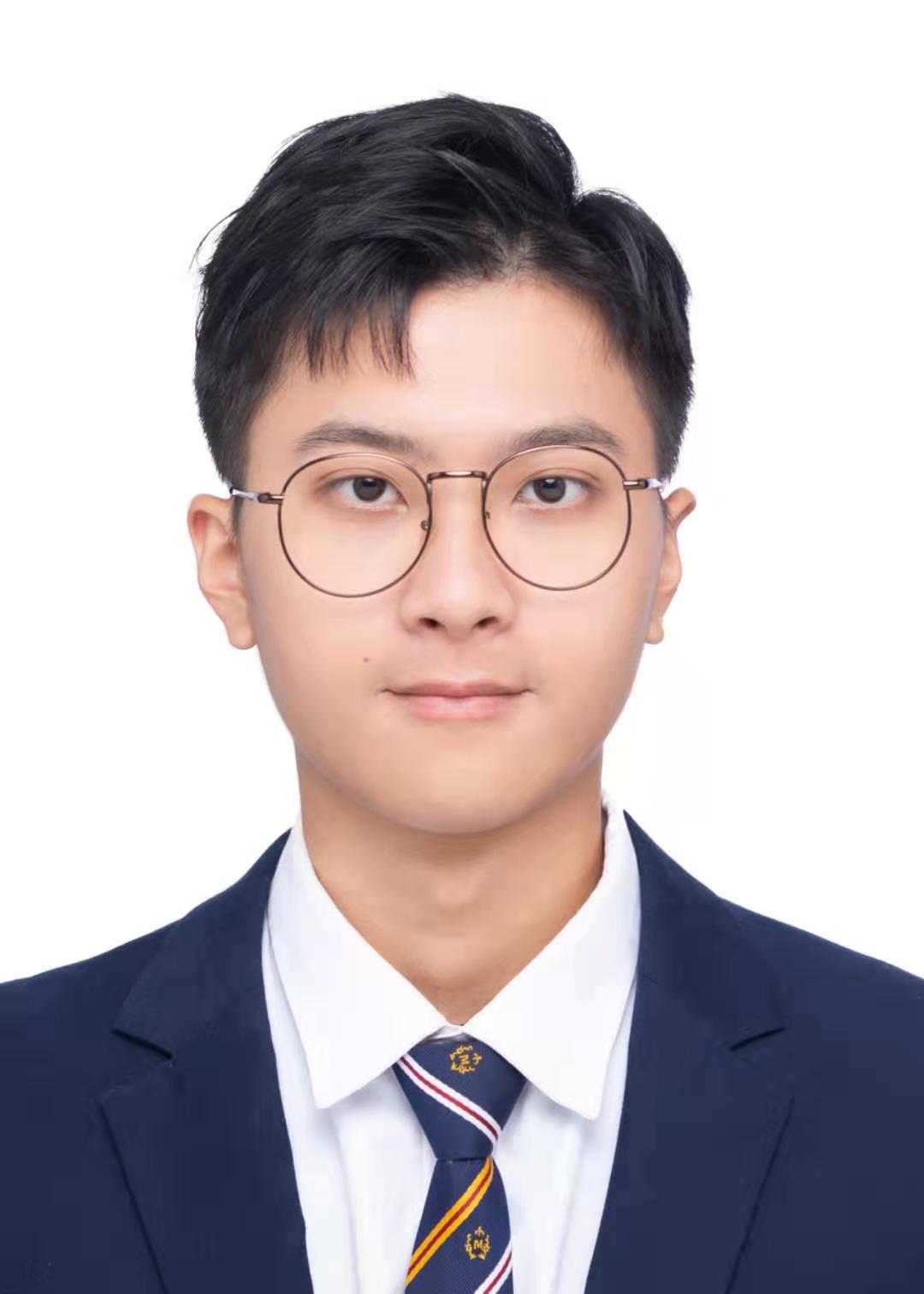}}]{Duo Wu} is currently a Ph.D. student at Tsinghua University. He received his M.Phil. degree and B.Eng. degree from The Chinese University of Hong Kong, Shenzhen in 2024 and Jinan University in 2022, respectively. He has broad  research interests in multimedia networking, reinforcement learning and large language models. He has published several first-author papers at top journals and conferences, including IEEE Transactions on Mobile Computing and ACM SIGCOMM.
\end{IEEEbiography}
\vspace{-1.5cm}
\begin{IEEEbiography}[{\includegraphics[width=1in,height=1.25in,clip,keepaspectratio]{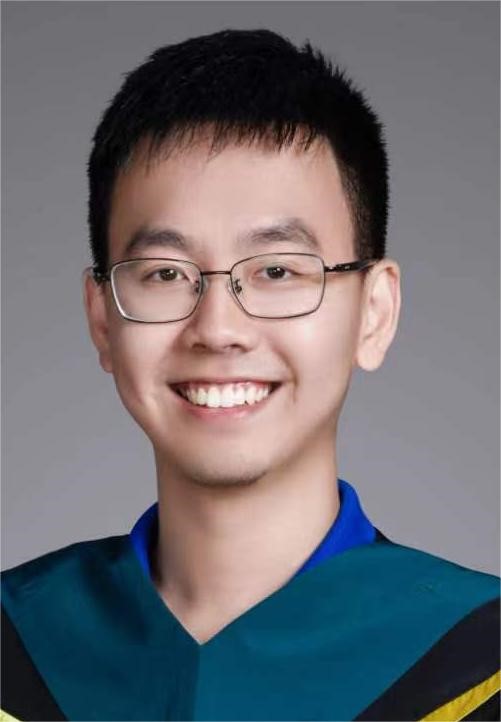}}]{Panlong Wu} received the B.Eng. degree from the Department of Electrical and Electronic Engineering, Southern University of Science and Technology in 2022. He is currently pursuing the Ph.D. degree in the School of Science and Engineering, The Chinese University of Hong Kong, Shenzhen. His current research interests include federated learning, foundation models and multimedia networking.
\end{IEEEbiography}
\vspace{-1.5cm}
\begin{IEEEbiography}[{\includegraphics[width=1in,height=1.25in,clip,keepaspectratio]{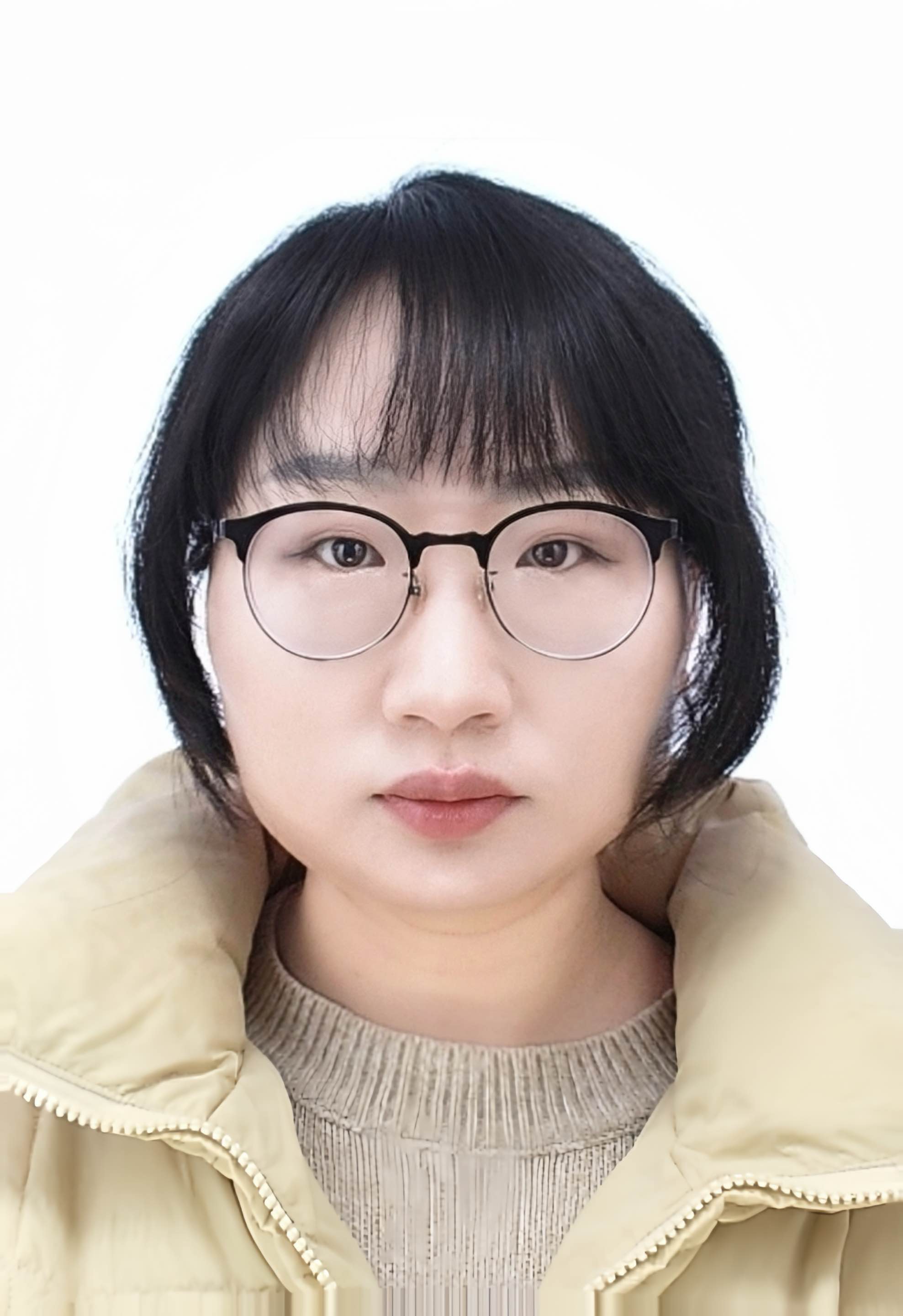}}]{Miao Zhang}
 (Student Member, IEEE) received her B.Eng. degree from Sichuan University in 2015, and her M.Eng. degree from Tsinghua University in 2018. She is currently a Ph.D. student at Simon Fraser University, British Columbia, Canada. Her research areas include cloud and edge computing, and multimedia systems and applications.
\end{IEEEbiography}
\vspace{-1.5cm}
\begin{IEEEbiography}[{\includegraphics[width=1in,height=1.25in,clip,keepaspectratio]{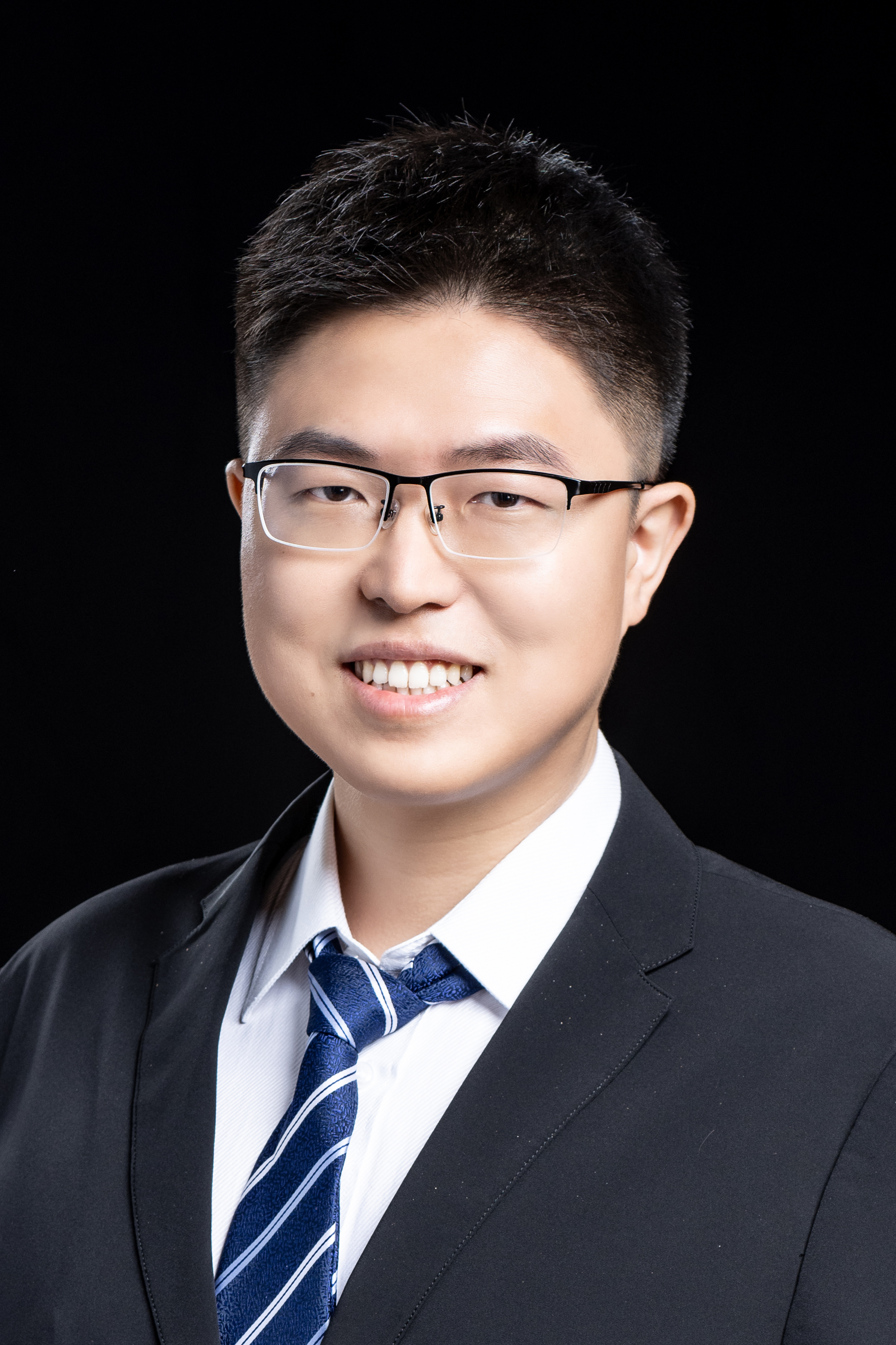}}]{Fangxin Wang}(S'15-M'20) is an assistant professor at The Chinese University of Hong Kong, Shenzhen (CUHKSZ). He received his Ph.D., M.Eng., and B.Eng. degree all in Computer Science and Technology from Simon Fraser University, Tsinghua University, and Beijing University of Posts and Telecommunications, respectively. Before joining CUHKSZ, he was a postdoctoral fellow at the University of British Columbia. Dr. Wang's research interests include Multimedia Systems and Applications, Cloud and Edge Computing, Deep Learning, and Distributed Networking and System. He leads the intelligent networking and multimedia lab (INML) at CUHKSZ. He has published more than 50 papers at top journal and conference, including INFOCOM, Multimedia, VR, ToN, TMC, IOTJ, etc. He was selected in the 8th Young Elite Scientist Sponsorship Program, CUHKSZ Presidential Young Scholar, and a recipient of SFU Dean's Convocation Medal for Academic Excellence. He serves as an associate editor of IEEE Transactions on Mobile Computing, TPC chair of IEEE Satellite 2023, TPC member of IWQoS, ICC, BigCom and reviewer of many top conference and journals, including INFOCOM, ToN, TMC, JSAC, etc.
\end{IEEEbiography}
% \vskip -2\baselineskip plus -1fil


%\newpage



%\vfill


\end{document}